\documentclass[a4paper,11pt]{article}

\usepackage[usenames,dvipsnames,table]{xcolor}
\usepackage[titletoc]{appendix}
\usepackage{mathtools,amssymb}
\usepackage[makeroom]{cancel}
\usepackage{bigints}
\usepackage{stackrel}
\usepackage{graphicx}
\usepackage{epstopdf}
\usepackage{mathrsfs}
\usepackage{enumitem}
\usepackage{float}
\usepackage{booktabs}
\usepackage[normalem]{ulem}  
\usepackage{longtable}
\usepackage{hyperref}

\DeclareMathAlphabet{\mathpzc}{OT1}{pzc}{m}{it}

 \allowdisplaybreaks
 \allowbreak

\newcommand{\lsim}{\mathrel{\hbox{\rlap{\lower .55ex
\hbox{$\sim$}} \kern-.3em \raise.4ex \hbox{$<$}}}}
\newcommand{\gsim}{\mathrel{\hbox{\rlap{\lower.55ex
\hbox{$\sim$}} \kern-.3em \raise.4ex \hbox{$<$}}}}

\begin{document}


\newcommand{\partiald}[2]{\dfrac{\partial #1}{\partial #2}}
\newcommand{\be}{\begin{equation}}
\newcommand{\ee}{\end{equation}}
\newcommand{\f}{\frac}
\newcommand{\s}{\sqrt}
\newcommand{\p}{\partial}
\newcommand{\lm}{\mathcal{L}}
\newcommand{\wm}{\mathcal{W}}
\newcommand{\om}{\mathcal{O}_{n}}
\newcommand{\bea}{\begin{eqnarray}}
\newcommand{\eea}{\end{eqnarray}}
\newcommand{\ba}{\begin{align}}
\newcommand{\ea}{\end{align}}
\newcommand{\ep}{\epsilon}

\def\gap#1{\vspace{#1 ex}}
\def\be{\begin{equation}}
\def\ee{\end{equation}}
\def\bal{\begin{array}{l}}
\def\ba#1{\begin{array}{#1}}  
\def\ea{\end{array}}
\def\bea{\begin{eqnarray}}
\def\eea{\end{eqnarray}}
\def\beas{\begin{eqnarray*}}
\def\eeas{\end{eqnarray*}}
\def\del{\partial}
\def\eq#1{(\ref{#1})}
\def\fig#1{Fig \ref{#1}} 
\def\re#1{{\bf #1}}
\def\bull{$\bullet$}
\def\nn{\nonumber}
\def\ub{\underbar}
\def\nl{\hfill\break}
\def\ni{\noindent}
\def\bibi{\bibitem}
\def\ket{\rangle}
\def\bra{\langle}
\def\vev#1{\langle #1 \rangle} 
\def\mattwo#1#2#3#4{\left(\begin{array}{cc}#1&#2\\#3&#4\end{array}\right)} 
\def\tgen#1{T^{#1}}
\def\half{\frac12}
\def\floor#1{{\lfloor #1 \rfloor}}
\def\ceil#1{{\lceil #1 \rceil}}

\def\mysec#1{\gap1\ni{\bf #1}\gap1}
\def\mycap#1{\begin{quote}{\footnotesize #1}\end{quote}}

\def\Red#1{{\color{red}#1}}

\def\Om{\Omega}
\def\a{\alpha}
\def\b{\beta}
\def\l{\lambda}
\def\g{\gamma}
\def\eps{\epsilon}
\def\vp{\varphi}
\def\bg{{\bar g}}

\def\lan{\langle}
\def\ran{\rangle}

\def\bit{\begin{item}}
\def\eit{\end{item}}
\def\benu{\begin{enumerate}}
\def\eenu{\end{enumerate}}

\def\tr{{\rm tr}}


\parindent=0pt
\parskip = 10pt

\def\a{\alpha}
\def\g{\gamma}
\def\G{\Gamma}
\def\b{\beta}
\def\d{\delta}
\def\D{\Delta}
\def\e{\epsilon}
\def\fb{{\bar f}}
\newcommand{\prop}{G_{\L}}
\newcommand{\propn}{G_{\L'}}
\def\k{\kappa}
\def\r{\rho}
\def\rvec{\vec{\rho}}
\def\l{\lambda}
\def\L{\Lambda}
\def\mO{\mathcal{O}}
\def\mbO{{\mathbb O}}
\def\P{\Phi}
\def\s{\sigma}
\def\t{\theta}
\def\z{\zeta}
\def\n{\eta}
\def\del{\partial}
\def\bt{\bar{\t}}
\def\x{{\rm x}} 
\def\scA{\mathcal{A}}
\def\scAb{\mathbb{A}}
\def\md{\mathsf{d}}
\def\mL{\mathcal{L}}
\def\mPz{\mathcal{P}_{\Delta z}}
\def\mQz{\mathcal{Q}_{\Delta z}}
\def\mRz{\mathcal{R}_{\Delta z}}
\def\mK{\mathcal{K}}
\def\Gcoff{\mathbb{G}}
\newcommand{\mD}{\mathcal{D}}
\def\fb{\bar{f}}
\def\gb{\bar{g}}
\def\tl{\tilde{\l}}
\def\pt{\tilde{\phi}}
\def\pb{\bar{\phi}}
\newcommand{\phiO}{\phi^{(0)}}


\newcommand*{\Cdot}[1][1.25]{%
  \mathpalette{\CdotAux{#1}}\cdot%
}
\newdimen\CdotAxis
\newcommand*{\CdotAux}[3]{%
  {%
    \settoheight\CdotAxis{$#2\vcenter{}$}%
    \sbox0{%
      \raisebox\CdotAxis{%
        \scalebox{#1}{%
          \raisebox{-\CdotAxis}{%
            $\mathsurround=0pt #2#3$%
          }%
        }%
      }%
    }%
    \dp0=0pt %
    \sbox2{$#2\bullet$}%
    \ifdim\ht2<\ht0 %
      \ht0=\ht2 %
    \fi
    \sbox2{$\mathsurround=0pt #2#3$}%
    \hbox to \wd2{\hss\usebox{0}\hss}%
  }%
}

\renewcommand{\arraystretch}{2.5}%
\renewcommand{\floatpagefraction}{.8}%

\def\bfnd{\bar{\mathfrak{f}}}
\def\bfd{\mathfrak{f}}
\def\fnd{\bar{f}}
\def\fd{f}
\def\bfndD{\bar{\mathpzc{f}}}
\def\bfdD{\mathpzc{f}}

\def\DCT{\hat{\mathcal{D}}_{ct}}

\def\Jc{\mathcal{J}}
\def\corrM{\mathfrak{M}}
\def\PermP{\mathcal{P}}
\def\afxd{\mathscr{A}^*_{ST}}

\hypersetup{pageanchor=false}
\begin{titlepage}
\begin{flushright}
TIFR/TH/16-27
\end{flushright}

\vspace{.4cm}
\begin{center}
\noindent{\Large \bf{Defining AdS/CFT at a finite radial cut-off}}\\
\vspace{1cm}
Gautam Mandal\footnote{mandal@theory.tifr.res.in}
and 
Pranjal Nayak\footnote{pranjal@theory.tifr.res.in}

\vspace{.5cm}
\centerline{{\it Department of Theoretical Physics}}
\centerline{{\it Tata Institute of Fundamental Research, Mumbai 400005, 
India.} }

\gap2

\today

\end{center}

\begin{abstract}

We define AdS/CFT at a finite radial cut-off, specifically in the
context of double trace perturbations, ${\mathbb O}_n$= ${\mathcal
  O}(x) (\partial^2)^{n} {\mathcal O}(x)$, with arbitrary powers
$n$. As well-known, the standard GKPW prescription, applied to a
finite radial cut-off, leads to contact terms in correlators. de Haro
et al \cite{deHaro:2000vlm} introduced bulk counterterms to remove
these. This prescription, however, yields additional terms in the
correlator corresponding to spurious double trace
deformations. Further, if we view the GKPW prescription coupled with
the prescription in \cite{deHaro:2000vlm}, in terms of a boundary
wavefunction, we find that it is incompatible with radial
Schr\"{o}dinger evolution (in the spirit of holographic Wilsonian
RG). We consider a more general wavefunction satisfying the
Schr\"{o}dinger equation, and find that generically such wavefunctions
generate both (a) double trace deformations and (b) contact
terms. However, we find that there exist special choices of these
wavefunctions, amounting to a new AdS/CFT prescription at a finite
cut-off, so that both (a) and (b) are removed and we obtain a pure
power law behaviour for the correlator. We compare these special
wavefunctions with a specific RG scheme in field theory.
We give a geometric interpretation of these wavefunctions; these
correspond to some specific smearing of boundary points in the Witten
diagrams. We present a comprehensive calculation of exact double-trace
beta-functions for all couplings ${\mathbb O}_n$ and match with a
holographic computation using the method described above. The matching
works with a mapping between the field theory and bulk couplings; such
a map is highly constrained because the beta-functions are quadratic
and exact on both sides. Our discussions include a generalization of
the standard double-trace Wilson-Fisher flow to the space of the
infinite number of couplings.

\end{abstract}

\gap1

\end{titlepage}
\hypersetup{pageanchor=true} \tableofcontents \thispagestyle{empty}
\enlargethispage{1000pt}
\pagebreak
\setcounter{page}{1}

\section{Introduction and Summary}

In AdS/CFT, conformal field theory partition function at a finite UV
cut-off ($\L$) is given by an AdS partition function at a finite
radial cut-off $z= \e= R_{AdS}^2/\L$. The latter quantity, of course,
needs a boundary condition. For example, the original GKPW prescription is a
Dirichlet boundary condition. It is well-known, however, that the bulk
path integral with this boundary condition leads to correlators with
contact terms some of which may diverge in the limit $\e\to 0$.
Following de Haro et al \cite{deHaro:2000vlm}, 
it is possible to add bulk counterterms to
remove these contact terms (completely or partially). With recent
insight from hWRG (holographic Wilsonian RG \cite{Heemskerk:2010hk,Faulkner:2010jy}), we will treat boundary conditions at $z=\e$ as a
wavefunction $\Psi_0[\phi_0, \e]$ (e.g.  Dirichlet b.c. is a
delta-function wavefunction). Some obvious questions which arise are

(1) What are the allowed boundary wavefunctionals (equivalently,
boundary conditions)?\\ 
(2) What does a choice of boundary condition/wavefunction in the
bulk path integral correspond to in the CFT?

The answer to question (1) is obvious from the discussions on hWRG. A
boundary wavefunction $\Psi_0[\phi_0, \e]$ is allowed provided its
$\e$-dependence follows the radial Schr\"{o}dinger equation $\del_\e
\Psi_0[\phi_0, \e]$ = $H_{rad}[ \phi_0, \del/\del \phi_0]\;
\Psi_0[\phi_0, \e]$. In the limit of $G_N \to 0$ (implicit in the
above equation), the Schr\"{o}dinger equation reduces to a Hamilton-Jacobi
equation for $S[\phi_0, \e]= \log \Psi_0[\phi_0, \e]$:
\[
\del_\e S =  H_{rad}[ \phi_0, \del S/\del \phi_0]
\]
For example, for a quadratic bulk action such as \eq{bulk-action}, the space
of allowed boundary wavefunctions $\Psi_0= e^S$ is given by the
\eq{a-b-c}, which we reproduce schematically as (here we suppress
the $\e$-dependent factors in ${\bf B, C}$)
\begin{align}
\Psi_0[\phi_0;\e] = \exp\kern-3pt\left[ -\f12\kern-3pt \int \kern-5pt
  \sqrt{\g_0} \left( {\bf A}(k, \e) \phi_0(k) \phi_0(-k) + 2 {\bf
    B}(k, \e) J(k) \phi_{0}(-k) + {\bf C}(k, \e_0) J(k) J(-k)\right)
  \right]
\end{align}
We will show below that the wavefunctional corresponding to GKPW 
\cite{Gubser:1998bc, Witten:1998qj}
boundary conditions, normally taken to represent the CFTs (Dirichlet
boundary condition for standard quantization and Neumann for alternative
quantization when the latter exists), correspond to a wavefunctional with a wrong $\e$-dependence when taken with the 
counterterms in \cite{deHaro:2000vlm}, as they do not satisfy the radial Schr\"{o}dinger equation.
This wavefunctional also leads to spurious double trace
deformations in the dual CFT. The correct wavefunctions which
represent the IR and UV CFT's (standard and alternative CFTs) are the
wavefunctions $\Psi_1^{0}$ and $\Psi_2^{0}$ described below (\autoref{eq:wavefunctional-correct-st} and \ref{eq:wavefunctional-correct-aq}, respectively).\footnote{$\Psi_2^0$, the wavefunctional corresponding to the UV fixed point has an interpretation of a unitary quantum field theory only inside the Klebanov-Witten window.}

A partial answer to question (2) appears in \cite{Witten:2001ua} where it is
shown that a subset of the above wavefunctions represents a CFT with
double-trace deformations (see Section \ref{sec:pres}). In the present
work, we will give a detailed and improved interpretation of the $A, B, C$
coefficients\footnote{$J$ will continue to represent the source for
  the single trace operator $\mO(x)$ dual to the bulk field
  $\phi$.}. In particular we will show that various choices of the
$A,B,C$ terms correspond to (i) double-trace deformations,
\begin{align}
S= S_{CFT} + \sum_{n=0}^\infty f_n  \int \mbO_n, \quad 
{\mathbb O}_n = \mO(x) (\del^2)^n \mO(x)
\label{all-double-trace}
\end{align}
and (ii) contact terms.  We have summarized the interpretation of these coefficients in \autoref{table:a-b-c}. One of
the main observations of our paper will be that there exist special
wavefunctions (with special choices of $A,B,C$) such that both (i) and (ii)
are absent and the correlators become pure power laws. Indeed, as mentioned
above, there
are just two such special choices $\Psi_1^0$ and $\Psi_2^0$ in the context
studied in this paper: one corresponds to the IR CFT (standard
quantization) without any deformations and the other corresponds to
the UV CFT (alternative quantization) without any deformation. In
Section 2 these will correspond to setting a quantity $\chi(k)$
(characterizing $A,B,C$, and hence the boundary wavefunctional) to zero or $\infty$.

We will show (in \autoref{sec:geo-interpret}) that the wavefunctions $\Psi_1^0$ and $\Psi_2^0$ have a simple geometric interpretation. Each of them corresponds to a
specific smearing of the boundary
points in Witten diagrams; as
mentioned above, the defining property of the above smearing is that
even when
the cut-off surface is moved inside, the resulting
correlators remain a power law. As an application of the above insight, we compute the Wilsonian
holographic beta-functions of the double-trace operators
and compare them with those obtained from direct
calculations in field theory. We find that the infinite number of
coupled beta-functions can be exactly mapped between field theory and holographic calculations. The existence of such a mapping is nontrivial since both the 
field theory 
and holographic beta-functions are exact and strictly quadratic.
The correct identification of the double trace deformations with
the boundary wavefunctionals plays here an essential role.

The organization of the paper is as follows:

In \autoref{sec:implications}, we discuss the allowed boundary conditions at finite
cut-off and arrive at the two wavefunctionals $\Psi_1^0$ and $\Psi_2^0$
which correctly represent the IR and UV CFTs respectively. In \autoref{sec:geo-interpret}, we discuss a geometric interpretation
of the wavefunctions $\Psi_1^0$ and $\Psi_2^0$. We show that each of
these represents introducing a specific kind of non-locality which
smears the boundary points in Witten diagrams in a particular way. Section \ref{sec:pres} presents the exact identification of the coefficients in a general boundary wavefunctional with coupling constants of double trace deformations
in Eq. \eq{all-double-trace} and the contact terms (a generic boundary wavefunction
represents both).  In \autoref{sec:hol-RG}, we use the above characterization of double trace
deformations  to compute the infinite series of
coupled beta-functions. In \autoref{sec:field-beta}, we present a detailed field
theory computation of these infinite series of beta-functions and
discuss the matching between the two results in \autoref{sec:rational-fraction}. The matching works with
a mapping between the FT and bulk couplings; such a map is highly
constrained because the beta-functions are quadratic and exact on both
sides. In \autoref{sec:discussions}, we discuss some outstanding problems. The details of most of the calculations have been reserved to appendices. \autoref{sec:notations} lists some of the notations for the double-trace couplings constants that are followed throughout the paper. \autoref{app:results} presents some mathematical results that are used in the field theory computations of \autoref{sec:field-beta}. All the exact holographic $\b$-function calculations are in \autoref{app:Eff-Act}, where we have also discussed the RG flow between the standard(IR) and alternative(UV) theories in \autoref{app-sec:Comments-Standard}. Some comparison with known results in large N $O(N)$ vector model is discussed in appendix \ref{app:wilson-fisher}. Lastly, \autoref{app:probe-HJ} presents some general discussion of large N limits, probe approximation in AdS geometry and applicability of Hamilton-Jacobi equations in general.

The details of various calculations in the paper are available at \href{https://arxiv.org/abs/1608.00411}{arXiv:1608.00411} as a Mathematica notebook named \emph{CalculationsFile.nb}.


\section{AdS/CFT at a finite radial cut-off: 
fixed points \label{sec:implications}}

In this section we will present a precise extension of the GKPW
prescription (\cite{Gubser:1998bc,Witten:1998qj}) to a finite cut-off.  
We will present the ideas in the
context of correlation functions of a single trace operator $\mO(x)$,
which is dual to a scalar field $\phi(z,x)$ in $d+1$ dimensional AdS
spacetime. The spacetime metric and the scalar action are given by\footnote{For simplicity we will consider a Euclidean metric. \label{ftnt:euclid}}
\begin{align}
ds^2 \equiv g_{MN} dX^M dX^N= \frac{dz^2 + dx_\mu dx_\mu}{z^2},
\label{ads-metric}
\\ 
S_b= \frac12 \int d^dx dz \sqrt{g} \left( (\del \phi)^2 + m^2 \phi^2\right)
\label{bulk-action}
\end{align}
The mass of the scalar field is chosen to satisfy the usual
mass-dimension relation 
\footnote{Later on, when we specifically discuss the
  Klebanov-Witten window $\nu \in (0,1)$, two distinct CFT duals can
  be found, corresponding to $\mO(x)$ having scaling dimensions
  $\Delta_\pm = d/2 \pm \nu$. For the new CFT, defined as `alternative
  quantization', the conformal dimension is
  $\Delta_-$.\label{ftnt:alternative}}
\begin{align}
\Delta = \Delta_+ \equiv d/2 + \nu,~~ \nu= \sqrt{d^2/4 + m^2
  R_{AdS}^2},
\label{mass-dim-st}
\end{align}
(we use units where $R_{AdS}=1$). For our purposes, the scalar action
is the only relevant part of the bulk action since we will work in the
``probe approximation'' in which the AdS metric $g_{MN}$ remains
unaltered (see Appendix \ref{app:probe-HJ}).

\subsection{Standard quantization}\label{sec:implications-fp}
As usual, we will call `standard quantization' the quantum theory
defined by the usual GKPW prescription, characterized by the
mass-dimension relation \eq{mass-dim-st}. Under special circumstances
we can define an `alternative quantization' (see footnote
\ref{ftnt:alternative} and more detailed discussions below). We will
denote various quantities associated with the `standard quantization'
by a subscript $+$ (e.g. $\Delta_+$) (and similarly those associated
with `alternative quantization' by a subscript $-$). 

Let us begin with the following {\it putative definition} of AdS/CFT
for standard quantization (GKPW)
\begin{align}\label{st-part-func}
& Z_+[J_k] = \langle \exp\left[ \int d^dk J_k \mO_{-k}
\right] \rangle_{+} 
=\int \mD \phi_0  \Psi_0[\phi_0;\e_0] \int \stackrel[z>\e_0]{}{\mD \phi}
e^{- S_b} 
\\
\label{st-GKPW}
&\Psi_0[\phi_0;\e_0]= \Psi_{\rm GKPW} \times \Psi_{ct}, 
\ \Psi_{\rm GKPW}=   \d\left(\phi_0(k) - \e_0^{d-\D_+} J(k)
\right),\nonumber \\
&\hspace{5cm} \Psi_{ct}= \exp\left(-\half\kern-8pt 
\bigintssss\limits_{z=\e_0}\kern-8 pt\sqrt{\g_0} \phi_k
            \DCT(k\e_0)\phi_{-k}\right)
\end{align}
Here $\g_0$ is the determinant of the induced metric $\g_{\mu\nu}$
at a radial cut-off $z=\e_0$. 

The $\d$-function above is equivalent to imposing the Dirichlet
boundary condition on the bulk field at $z=0$, where the boundary
value of the bulk field is related to the source, $J(k)$, of the dual
field theory operator $\mO(k)$ with some appropriate
renormalization. In addition to the original $\d$-function of GKPW, we
have also included the counter-terms denoted by $\DCT(k\e_0)$
conventionally introduced to ensure finiteness of the bulk partition
function in the $\ep \to 0$ limit \cite{deHaro:2000vlm} (see also
\cite{Klebanov:1999tb, Balasubramanian:1999re}; these counterterms can
also be motivated from the requirement of a well-defined variational
principle, cf.  \eq{variational} below). Expanded to several orders in
$(k\e_0)^2$, it reads (\cite{deHaro:2000vlm} gives the first two
terms; the expansion can be worked out to arbitrary orders with the
help of the Mathematica notebook {\it CalculationsFile.nb} in
\cite{MN:2016mathematica})
\begin{align}
\DCT(\e_0 k) &= \D_- - \frac{1}{2(\nu-1)}(k \e_0)^2  + 
\frac{1}{8 (\nu-2)(\nu-1)^2} (k \e_0)^4 + \cdots 
\label{contact}
\end{align}
As we will shortly see below, the precise choice of
counterterms is determined by demanding conformal invariance.

We will now show that the definition of AdS/CFT \eq{st-part-func}
using the wavefunctional \eq{st-GKPW} needs improvement in the sense that
the wavefunctional has a wrong dependence on $\e_0$. Of course, one
could take the viewpoint that it is meant to be valid only for a fixed
$\e_0$ and at other values the wavefunctional is different. It is not
clear what that special value is; one possibility is $\e_0=0$,
however, it is hardly clear how to take this limit in \eq{st-GKPW}. If
one does go ahead with this viewpoint and computes the RHS of
\eq{st-part-func} at some special value of $\e_0$, it {\sl does not
  give the right results expected in the CFT}, rather the correlators
computed from it are of the form \eq{pert-corr-ir} obtained from a
regulated field theory perturbed by double trace operators.
While, in some sense, these correlators do limit to those expected from conformal symmetry, strictly speaking, these can't be interpreted as coming from an exact conformal field theory through Wilsonian philosophy.

We therefore demand that the wavefunctional must be specified such
that it at least has the correct dependence on $\e_0$. We now discuss
the general class of such wavefunctions.

\paragraph{The space of allowed wavefunctionals:}

The general form of the wavefunctional $\Psi_0[\phi_0, \e_0]$, in
particular the dependence on $\e_0$, can be inferred from the fact
that it must satisfy the radial Schr\"{o}dinger equation, which, in the
case of a bulk theory with a free massive scalar without gravitational
back reaction, takes the form
  \begin{align}
    -\del_{\e_0} \Psi[\psi_0;\e_0] &= \hat{H}_{rad} \Psi[\psi_0;\e_0] \label{eq:bulk-evol}, ~\text{where,}\\
    \hat{H}_{rad} = \int d^dx \ \hat{\mathcal{H}}_{rad} &= \frac{1}{2} \left( \int d^d k \ \frac{1}{z^{1-d}} \hat{\Pi}_k \hat{\Pi}_{-k} + z^{-1-d} \left(z^2 k^2+m^2 \right) \hat{\phi}_k \hat{\phi}_{-k} \right)  
\text{and, } \hat{\Pi} \equiv i \frac{\d}{\d \phi} \nonumber
  \end{align}
The general solution for the wavefunctional is of the following
quadratic form in the bulk field $\phi_0$, of the form \footnote{The
  explicit $\e_0$-dependent factors in front of $B$ and $C$ are chosen
  so that the parameters $A,B,$ and $C$ in the wavefunctional are
  dimensionless (note our choice of units where $R_{AdS}=1$).  The
  form of the wavefunction can also be argued based on explicit
  integration of the near boundary degrees of freedom in the bulk
  action, as is done in \cite{Heemskerk:2010hk, Faulkner:2010jy,
    Elander:2011vh}, and also in \autoref{app:Eff-Act}. Without any interactions, the wavefunctional obtained by integrating out degrees of freedom between $z=0$ and some $z=\e_0$ can only be quadratic.}
\begin{align}\label{a-b-c}
\Psi_0[\phi_0;\e_0]  =  \exp\Bigg[ -\f12 \int\limits_{z=\e_0} \hspace{-7pt} d^dk 
\sqrt{\g_0} \Big( A(k, \e_0) \phi_k \phi_{-k} + 2 \e_0^{d-\D_+} B(k, \e_0) J_k 
\phi_{-k} \nonumber \\
+ \e_0^{2(d-\D_+)} C(k, \e_0) J_k J_{-k}\Big) \Bigg]
\end{align}
Eqn. \eq{eq:bulk-evol}, computed in the Hamilton-Jacobi
approximation \cite{Heemskerk:2010hk, Faulkner:2010jy} gives 
\footnote{In this particular quadratic case, Hamilton-Jacobi approximation is equivalent to exact Schr\"{o}dinger equations. The second and third equations of \eq{abc-dot} are slightly
  different from the corresponding equations in
  \cite{Heemskerk:2010hk, Faulkner:2010jy} due to the fact that the
  their $B,C$ are dimensionful.}
\begin{align}\label{abc-dot}
  	\dot A= - (A- \D_+)(A- \D_-) + (k\e)^2, \quad   \dot B = \D_+ \; B- A\; B, \quad \dot C = (2\D_+-d)\; C-B^2
  \end{align}
here, $\dot{X}$ denotes, $\e_0\del_{\e_0}X$. The general closed form solution for $A(k,\e_0)$ is,
\begin{align}\label{eq:A-gen-sol}
	A(k,\e)=\frac{\chi(k) \ \Big( (\f d2+ \nu ) I_{-\nu }(k \e ) + k \e_0  I_{-\nu -1}(k \e ) \Big)+ (-1)^{ \nu } \frac{\Gamma (\nu +1)}{\Gamma (1-\nu )} \Big( (\f d2- \nu ) I_{\nu }(k \e )+ k \e  I_{\nu -1}(k \e ) \Big)}{ \chi(k) \ I_{-\nu }(k \e )+ (-1)^{\nu } \frac{\Gamma (\nu +1)}{\Gamma (1-\nu )} I_{\nu }(k \e )} \nonumber \\[10pt]
	 = \frac{2^{\nu } \chi(k)  \left( (d-2 \nu ) + (k \e) ^2  \ \frac{(d-2 \nu +4)}{4 (1-\nu )} + \ldots \right) +\left(-\frac{1}{2}\right)^{\nu } (k \e)^{2 \nu } \left( (d+2 \nu ) + (k \e)^2 \frac{ (d+2 \nu +4)}{4 (\nu +1)} + \ldots \right) }{ 2^{\nu +1} \chi(k)  \left( 1 + (k \e)^2\frac{ 1}{4 (1-\nu )}+ \ldots \right)+2 \left(-\frac{1}{2}\right)^{\nu } (k \e)^{2 \nu } \left( 1 + (k \e)^2 \frac{1}{2 (4 (\nu +1))}+ \ldots \right)}
\end{align}
Here, $\chi(k)$ is a constant of integration, fixed by solving with a boundary condition at some cut-off $z=\e_0$. Note that the above solution in the series form has two independent series, a series in integer powers of $(k\e_0)$ and another series in powers of $(k\e_0)^{2\nu}$. We will show later that the series corresponding to $(k\e_0)^{2\nu}$ contains information about the double trace deformations around the fixed point.\\
Similar solutions exist for $B(k, \e_0)$ and $C(k, \e_0)$. 

\paragraph{Wavefunctional satisfying exact scaling} In general, the partition function can be computed by integrating out the bulk fields exactly,
\begin{align}\label{exact-partition-function}
	Z[J_k] = \exp \left[  -\half \int d^dk \ J_k J_{-k} \e_0^{d-2\D_+} \left( C(k,\e_0) - \frac{B^2(k,\e_0)}{k\e_0 \frac{K_{\nu-1}(k\e_0)}{K_\nu(k\e_0)} - \D_- + A(k,\e_0)} \right) \right]
\end{align}
where, $K_\nu(k\e_0)$ are the modified Bessel functions of second kind. There are two special choices of $\chi(k)$ above, i.e. $\chi(k)=0$, or $\infty$, for which the partition function in \eq{exact-partition-function} becomes exactly that of a conformal theory.\footnote{\label{foot:dot-explanation} What we really mean here is that the partition function computed above doesn't explicitly depend on the cut-off $\e_0$, thus obeying the correct scaling laws corresponding to the dual field theory operator $\mO$. This is also the reason to claim that such a wavefunctional can be understood as being generated by integrating out the degrees of freedom between $z=0$ and $z=\e_0$ in the bulk theory that is exactly dual to the conformal field theory, the limiting action given by \eqref{st-GKPW}.} 
\vspace{5pt}
\begin{figure}[!ht]
\begin{center}
\includegraphics[scale=.25]{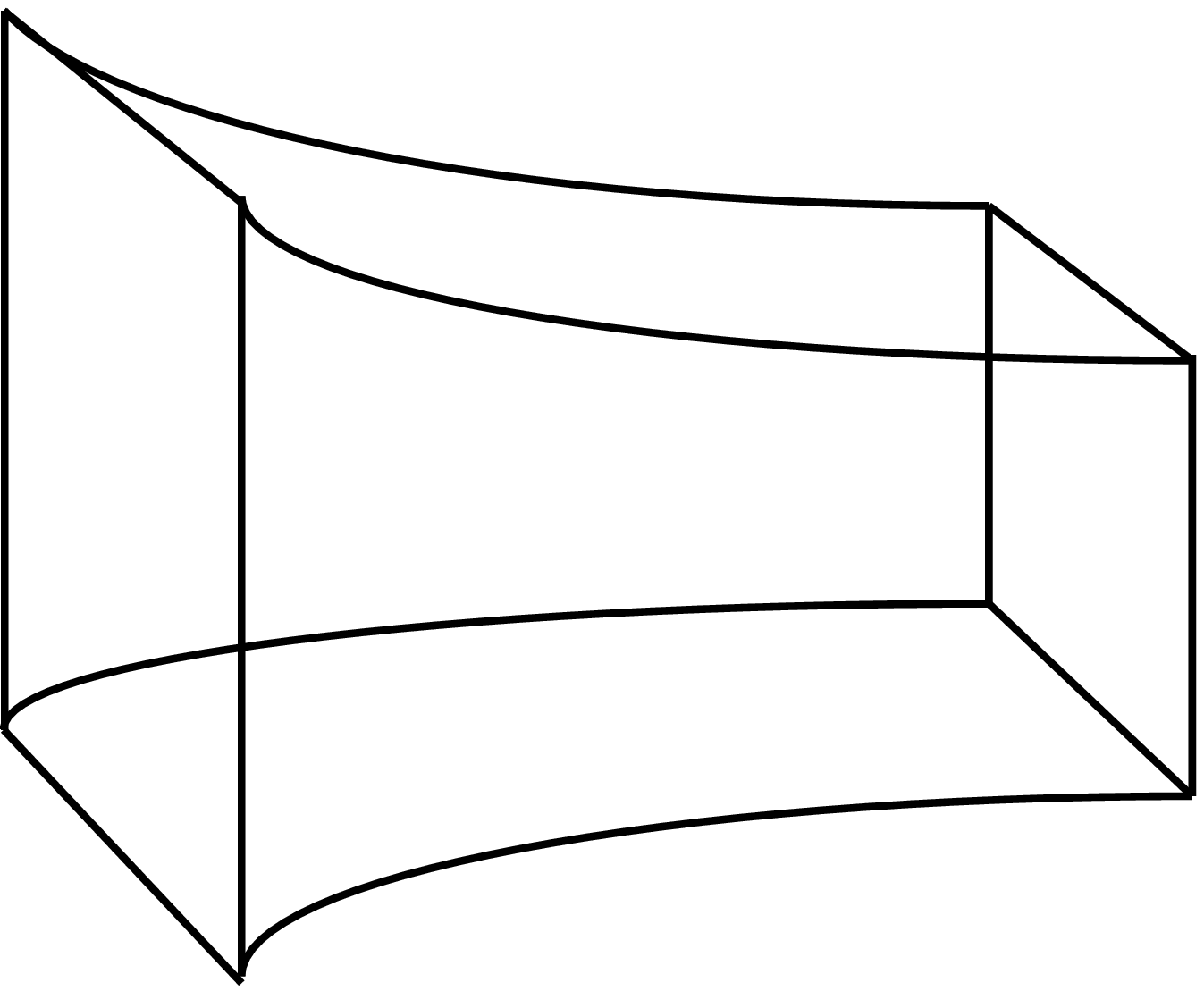}\hspace{3cm}
\includegraphics[scale=.25]{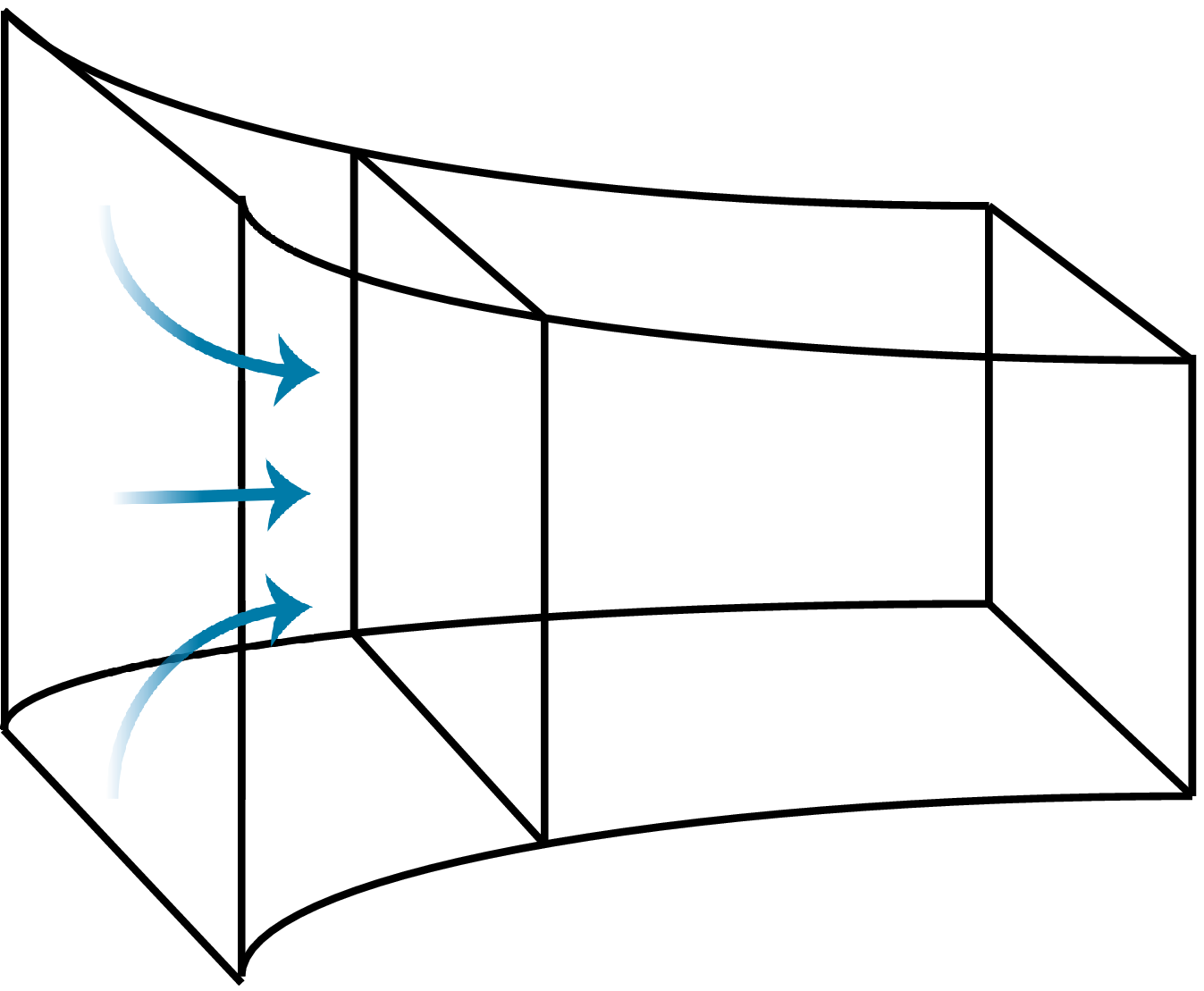}
\caption{The wavefunctional with the coefficients $A^*, B^*, C^*$ gives the correct effective description of the continuum theory, which was obtained by the $\e_0=0$ bulk action. This wavefunctional is effectively obtained by the integration of near boundary degrees of freedom in AdS.}
\end{center}
\label{fig:fixed-pt-bulk-integration}
\end{figure}

To the leading order in $k\e_0$, the solution for these particular choices of the wavefunctionals are $A=\D_+ \text{ or } \D_-$, as can also be seen from the leading order truncation of \eq{eq:A-gen-sol}. Let us consider the solution with $A=\D_+$. In this case, $B$-evolution equation is identically satisfied, and the value of $B$ is fixed by the boundary value enforced by \eq{st-GKPW} (to leading order, in continuum limit) to $B=-2\nu$. Finally, this fixes $C=2\nu$, and the wavefunctional is given by,
\begin{align}\label{eq:leading-Psi}
\Psi_0[\phi_0;\e_0]  &\sim  \exp\Bigg[ -\f12 \int\limits_{z=\e_0} \hspace{-7pt} d^dk 
\sqrt{\g_0} \Big( \D_+ \ \phi_k \phi_{-k} - 4\nu \e_0^{d-\D_+} J_k \phi_{-k} + 2\nu \e_0^{2(d-\D_+)} J_k J_{-k}\Big) \Bigg] \nonumber \\
& \sim \exp\Bigg[ -\f12 \times (2\nu) \int\limits_{z=\e_0} \hspace{-7pt} d^dk 
\sqrt{\g_0} \Big( \phi_k - \e_0^{d-\D_+} J_k \Big)_k \Big( \phi_k - \e_0^{d-\D_+} J_k \Big)_{-k} \Bigg] \nonumber \\
& \hspace{5cm} \times \exp \left[ -\f12 \D_- \int\limits_{z=\e_0} \sqrt{\g_0} \phi_k \phi_{-k}\right] \ \footnotemark
\end{align}
\footnotetext{`$\sim$' signifies that the subleading terms have not been included.}%
Note that with $A= \D_+, B= - 2\nu, C= 2\nu$, we have an appropriately 
regulated, and correct form of the wavefunctional \eq{st-GKPW}.

The solution with the sub-leading corrections can be found to arbitrary order in $(k\e_0)$ and are given by,
\begin{subequations}\label{eq:ABC-series-sol-st}
	\begin{align}
		A_{ST}^*(k\e_0) &= \D_+ + \frac{1}{2 (1+\nu)} (k\e_0)^2 - \frac{1}{8 (2+\nu) (1+\nu)^2} (k\e_0)^4 + \cdots \label{eq:asol-series}\\
		&= \DCT(k\e_0) + 2\nu \left( 1 - \frac{1}{2 \left( 1 - \nu ^2 \right)} (k\e_0)^2 + \frac{ \left( 5 + \nu ^2 \right)}{8 \left( 4 - \nu ^2 \right) \left( 1 - \nu ^2 \right)^2} (k\e_0)^4 + \cdots \right)\\
		&=  \DCT(k\e_0) + 1/\afxd\\[5pt]
		\afxd \cdot B_{ST}^*(k\e_0) &= - \left( 1+\frac{1}{4(1 -\nu )} (k\e_0)^2 + \frac{1}{32 (1-\nu)(2-\nu)} (k\e_0)^4 + \cdots \right)\label{eq:bsol-series}\\[5pt]
		\afxd \cdot C_{ST}^*(k\e_0) &= 1 + \frac{1}{2-2 \nu } (k\e_0)^2 + \frac{(3 - 2 \nu)}{16 (2 - \nu) (1 - \nu)^2} (k\e_0)^4+ \cdots \label{eq:csol-series}
	\end{align}
\end{subequations}
where it can be checked that $\left( \afxd \cdot B_{ST}^*(k\e_0)  \right)^2 = \afxd \cdot C_{ST}^*(k\e_0) $. So the wavefunctional at the finite cut-off is,
\begin{align}\label{eq:wavefunctional-correct-st}
	\Psi_1^{0}[\phi_0;\e_0] &= \exp\left[ -\half \hspace{-5pt}\int\limits_{z=\e_0} \hspace{-6pt}d^dk \sqrt{\g_0} \dfrac{ \left( \phi + \afxd \cdot B_{ST}^*(k\e_0) \ \e_0^{d-\D_+} J \right)_k \left( \phi + \afxd \cdot B_{ST}^*(k\e_0) \ \e_0^{d-\D_+} J \right)_{-k} }{\afxd(k\e_0)} \right. \nonumber \\
	&\hspace{5cm} - \half \int_{z=\e_0} \sqrt{\g_0} \phi_k \DCT(k\e_0) \phi_{-k} \Bigg]
\end{align}
and $\afxd$ is just the shorthand for the series,
\[
	\frac{1}{\afxd} = 2\nu \left( 1 - \frac{1}{2 \left( 1 - \nu ^2 \right)} (k\e_0)^2 - \frac{ \left( 5 + \nu ^2 \right)}{8 \left( 4 - \nu ^2 \right) \left( 1 - \nu ^2 \right)^2} (k\e_0)^4+ \cdots \right)
\]
We note here that not only does the $\d$-function in the standard quantization in AdS/CFT correspondence gets regulated at finite cut-off, even the source, $J$, for the dual field theory operator, $\mO$ gets renormalized. The wavefunction renormalization at the finite cut-off is given by $Z^{-1}_{J} = \left(- \afxd(k\e_0) \cdot B_{ST}^*(k\e_0) \right)^{-1} = Z_\mO$.\footnote{We define $\mO^{(\e)} = Z_\mO \cdot \mO^{(0)}$ and $J^{(\e)} = Z_{J} \cdot J^{(0)}$. Alternatively, we emphasize that the correct way to identify the source is through \eq{eq:wavefunctional-correct-st}, without any mention of wavefunction renormalization.}

With such a choice of wavefunction, the RHS and consequently the LHS
of \eq{st-part-func} will actually be {\sl independent} of the cut-off
parameter $\e_0$ (see \eq{eq:exact-finite-partition-st} below)!  Thus, although the holographic calculation appears to be done at a finite
radial cut-off $z=\e_0$, the functional integral is actually independent
of the cut-off. We will see below that the correlators computed
from this prescription exhibit a pure power law behaviour.

\paragraph{$\mO(k)\mO(-k)$ correlator}We compute the correlators with the new prescription for AdS/CFT at finite radial cut-off with the inclusion of the boundary wavefunctional \eq{eq:wavefunctional-correct-st} by integrating out the bulk fields $\phi$. The exact partition function becomes,
\begin{align}\label{eq:exact-finite-partition-st}
	Z_+[J_k] = \exp\left[ -\half \int d^dk \ J_k \left( k^{2\nu} \ \frac{2^{1-2\nu }  \Gamma (1-\nu )}{\Gamma (\nu )} \right) J_{-k} \right]
\end{align}
This is the exact partition function to all orders with the correct solutions of $\afxd,B^*,C^* $.\footnote{We have checked it to the sixth order in $k\e$ expansion, but with the inclusion  of the exact solutions for $\afxd,B^*,C^* $ this will hold true to all orders.} Thus the connected two point function for the boundary operator is,
\begin{equation}\label{eq:exact-2-pt-finite-st}
	\lan \mO(k) \mO(-k) \ran_+ = k^{2\nu} \ \frac{2^{1-2\nu }  \Gamma (1-\nu )}{\Gamma (\nu )} 
\end{equation}
This is the correct 2-point function as governed by conformal symmetry. If we follow the Wilsonian principles of integrating out the degrees of freedom such that all the physical observables remain invariant, then this is the wavefunctional that we will obtain from \eq{st-GKPW}. This result is slightly surprising because it tells us that it is possible to define AdS/CFT correspondence with a finite bulk cut-off, such that we still describe the field theory in the continuum limit. Alternatively, from the conventional renormalization point of view, in the field theory this is analogous to finding out all the correct counter-terms and/or vacuum energy terms that make the partition function at a finite cut-off exactly conformally invariant. This view point is discussed in detail in \autoref{sub-sec:choice-of-regulation-bulk}.

\paragraph{Correlator for a regulated field theory}
Since we want to find a bulk dual to field theory that is regulated at short distances (\autoref{sec:field-beta}), we want to introduce an explicit cut-off dependence in our correlator/partition function which replicates the regulation-dependence in the field theory (see \autoref{sub-sec:ft-corr-mom-sp}). A position space regulated correlator, \eqref{regulator}, in momentum space is given by \eqref{th-regulated-corr-mom}.
To include a similar regulation in the bulk calculation, we need to include an extra contact term piece in our bulk action,
\begin{equation}\label{eq:Sextra-st}
	S_{extra} = \half \int d^dk \ \e_0^{-d + 2(d-\D_+)} \d C(k\e_0) J_k J_{-k}
\end{equation}
which modifies the correlator \eq{eq:exact-2-pt-finite-st} to,
\begin{equation}\label{eq:counter-2pt-st}
	\lan \mO(k) \mO(-k) \ran_+ = k^{2\nu} \ \frac{2^{1-2\nu }  \Gamma (1-\nu )}{\Gamma (\nu )} + \e_0^{-2\nu} \ \d C(k\e_0)
\end{equation}
One could argue that any perturbation away from the fixed point could ideally be achieved by changing any of $A,B$ or $C$ away from the fixed point values, $A^*,B^*,C^*$. But as we will see in \autoref{sec:pres}, each of these coefficients have a different field theory interpretation of double-trace perturbation, wavefunction renormalization and contact terms in the correlators/partition function, respectively. So the change of each one of them contributes in a different manner to the observables like correlators of the theory.

\begin{table}[!ht]
\begin{center}
\begin{tabular}{c|c|c}
\toprule
$A(k\e)$&
$B(k\e)$&
$C(k\e)$\\
\midrule
Double-trace deformation&Wavefunction renormalization& 
Contact terms\\
\bottomrule
\end{tabular}
\caption{Interpretation of different coefficients in wavefunctional \eq{a-b-c} away from the fixed point values, $A^*,B^*,C^*$. This interpretation is slightly heuristic and the exact relations are given in \autoref{sec:pres}.}
\label{table:a-b-c}
\end{center}
\end{table}

We have studied the RG flows of theories regulated in this fashion in field theory and we will do a parallel calculation in the bulk. But before that we also establish the AdS/CFT duality at a finite cut-off in alternative quantization.

\subsection{Alternative Quantization\label{sec:alternate}}
In Klebanov-Witten window $\nu= \Delta_+ - d/2$ $ \in (0,1)$
\cite{Klebanov:1999tb} the bulk gravitational theory is dual to two
different quantum field theories in the boundary which are related to
each other through Legendre transform. Thus, the generating function
of one quantum field theory is the $1PI$ effective action of the other
and vice versa, with the distinction that $1PI$ effective action is
itself a local action for such theories.

Alternative fixed point can be understood as a UV completion of the standard IR theory within the Klebanov-Witten window by analysing the flow equations \eq{abc-dot}.\footnote{It is the solution corresponding to $\chi\to\infty$ in \eq{eq:A-gen-sol}, with the corresponding solutions for $B(k,\e_0)$ and $C(k,\e_0)$.} However, we treat this as a stand-alone prescription to begin with, and will connect them using the flow in double trace couplings in Appendix \ref{app-sec:Comments-Standard}. The usual AdS/CFT prescription for the alternative quantization is given by,
    \begin{align}\label{eq:Aq-original}
	Z_-[J_k] &= \langle \exp \int d^dk J_k \mO_{-k} \rangle_-
        \nonumber\\ &= \int\limits_{z\ge \e_0} \mD \phi \exp \left[ -S_b
          - \lim_{\e_0 \to 0} \left( \int\limits_{z=\e_0} d^dk \sqrt{\g_0} \e_0^{d-\D_-} \phi_k
          J_{-k} + \half \int\limits_{z=\e_0} d^dx \sqrt{\g_0} \phi_k
          \DCT(\e_0 k) \phi_{-k} \right)\right]
    \end{align}
The boundary part of the action, which is also the wavefunctional $\Psi[\phi_0]$, in the above equation is such that the variation principle imposes a modified Neumann condition on the boundary $z=\e_0 \to 0$. This relates the normalizable part of the classical solution for $\pi$ (conjugate momentum to the bulk field $\phi$) to the source, $J$ for the dual field theory operator $\mO$, which now has the conformal dimension $\D_- = d/2-\nu$, \cite{Klebanov:1999tb, Aharony:2015afa}. In this case, the wavefunctional can be generalized to a finite cut-off without any ambiguity. Evolution equations for alternative quantization in terms of $A,B,C$ are ($B,C$ equations are modified due to difference in normalization of the sources with respect to the bulk field $\phi$),
\begin{align}\label{eq:abc-dot-alt}
	\dot A= - (A- \D_+)(A- \D_-) + (k\e_0)^2, \quad   \dot B = \D_- \; B- A\; B, \quad \dot C = (2\D_--d)\; C-B^2
\end{align}
It can be checked immediately that $\DCT$ given by \cite{deHaro:2000vlm} is identically a stationary point for $A$. At the leading order in $k\e_0$, $A=\D_-$ and $B=1$, and $B$ equation is identically satisfied. However, in the limiting prescription of \eqref{eq:Aq-original}, we don't have any $C$, which clearly is not a stationary point. As we saw in our concluding discussion in the previous section, $C$ terms are quadratic in the sources $J_k$ and add contact terms to the bulk action and the $\mO$ correlators, hence are interpretable as choice of regulation scheme at finite cut-off. We modify the wavefunctional in \eqref{eq:Aq-original} to include such terms and demand that this be at a fixed point as we did for standard quantization. We see later that inclusion of such a term makes the alternative theory the exact Legendre transform of the standard theory along with all the counter-terms in both the theories. Solving for the stationary point of $C$ to the leading order, the wavefunctional becomes,
\begin{align}\label{eq:leading-Psi-alt}
	\Psi_0[\phi_0;\e_0]  &\sim  \exp\Bigg[ -\f12 \int\limits_{z=\e_0} \hspace{-7pt} d^dk \sqrt{\g_0} \Big( \D_- \ \phi_k \phi_{-k} + 2 \e_0^{d-\D_-} J_k \phi_{-k} - \frac{1}{2\nu} \e_0^{2(d-\D_-)} J_k J_{-k}\Big) \Bigg] 
\end{align}
with the inclusion of the corrections in $k\e_0$, the wavefunctional becomes,
\begin{align}\label{eq:wavefunctional-correct-aq}
	\Psi_2^{0}[\phi_0;\e_0] &= \exp \Bigg[ -\half \int\limits_{z=\e_0}d^dk \sqrt{\g_0} \left( \phi_k \DCT(k\e_0) \phi_{-k} + 2 \e_0^{d-\D_-} B_{AQ}^*(k\e_0) \phi_k J_{-k} \right.  \nonumber \\
	& \hspace{5cm} \left.  + \e_0^{2(d-\D_-)} C_{AQ}^*(k\e_0) J_k J_{-k}\right) \Bigg]
\end{align}
where,
\begin{subequations} \label{eq:ABC-series-sol-aq}
\begin{align}
	B_{AQ}^*(k\e_0) &= 1 - \frac{1}{4 (1-\nu)} (k\e_0)^2 + \frac{ (3-\nu)}{32 (2-\nu) (1-\nu)^2} (k\e_0)^4 + \cdots \label{eq:bsol-series-aq}\\[5pt]
	C_{AQ}^*(k\e_0) &= -\frac{1}{2 \nu } +\frac{1}{4(1- \nu ^2)}  (k\e_0)^2- \frac{(5-2 \nu )}{32 (1-\nu)^2 \left(4-\nu ^2\right)} (k\e_0)^4 + \cdots \label{eq:csol-series-aq}
\end{align}\end{subequations}
It is interesting to note that, $B^*_{AQ}(k\e_0) = -1/(\afxd \cdot B^*_{ST}(k\e_0))$ and $C^*_{AQ}(k\e_0) = -1/(\afxd \cdot {B^*_{ST}}^2)$. This shows that the alternative theory given by the wavefunctional \eqref{eq:wavefunctional-correct-aq} is exactly the Legendre transform of the standard theory defined by the wavefunctional \eqref{eq:wavefunctional-correct-st} at cut-off $z=\e_0$.

\paragraph{$\mO(k)\mO(-k)$ correlator} The partition function and the correlator computation follows similar to that in standard quantization and can be computed exactly by using the wavefunctional \eq{eq:wavefunctional-correct-aq}, and integrating out the $\phi$ fields in the bulk,
\begin{align}\label{eq:exact-finite-partition-aq}
	Z_+[J_k] = \exp\left[ -\half \int d^dk \ J_k \left( -k^{-2 \nu } \ \frac{2^{2\nu -1}  \Gamma (\nu )}{\Gamma (1-\nu )} \right) J_{-k} \right]
\end{align}
Again, this is the exact correlator to all orders with the correct solutions of $A^*,B^*,C^*$. Thus the connected two point function for the boundary operator is,
\begin{equation}\label{eq:exact-2-pt-finite-aq}
	\lan \mO(k) \mO(-k) \ran_- = -k^{-2 \nu } \ \frac{2^{2\nu -1}  \Gamma (\nu )}{\Gamma (1-\nu )}
\end{equation}
This is the correct 2-point function as governed by conformal symmetry for a continuum theory around the UV-fixed point.

\paragraph{Correlator for a regulated field theory} Following the discussion in previous subsection, we can study a regulated field theory by including an extra piece in the wavefunctional, \eqref{eq:wavefunctional-correct-aq},
\begin{equation}\label{eq:Sextra-aq}
	S_{extra} = \half \int d^dk \ \e_0^{-d + 2(d-\D_-)} \d C(k\e_0) J_k J_{-k}
\end{equation}
which again modifies the correlator above to,
\begin{equation}\label{eq:counter-2pt-aq}
	\lan \mO(k) \mO(-k) \ran_- = -k^{-2 \nu } \ \frac{2^{2\nu -1}  \Gamma (\nu )}{\Gamma (1-\nu )} + \e_0^{2\nu} \d C(k\e_0)
\end{equation}

\subsection{Choice of regulation scheme and comparison with field theory}\label{sub-sec:choice-of-regulation-bulk}
In a dual field theory calculation, Wilsonian principles demand that under integration of degrees of freedom in a field theory, all physical observables remain unchanged. This gives us an effective description of the same theory with reduced degrees of freedom. In particular, if we start with a continuum quantum field theory and integrate out the UV degrees of freedom (either in position or momentum space), then the correlation functions computed using the new effective Lagrangian are the same as that of the continuum theory. In a continuum conformal field theory in which the correlation functions of the primary operators obey the scaling laws, an effective description with integration of certain degrees of freedom will reproduce the same power law correlators. However, a particular choice of regulation scheme in the field theory changes the short-distance/UV behaviour of the correlators (e.g. \eqref{regulator}) by an addition of certain counter-terms in the momentum space (\autoref{gen-regulatedcorr}). For example, for the $\Theta$-function regulated theory this choice corresponds to, (see \eqref{th-regulated-corr-mom}),
\begin{equation}\begin{aligned}
	c_0 = \pm \frac{2 \pi ^{\frac{d-1}{2}} }{\nu \  \Gamma \left(\dfrac{d-1}{2}\right)} , \hspace{0.25cm}
	c_1 = \frac{ \pi ^{\frac{d-1}{2}} }{3 (\nu+1)  \Gamma \left(\dfrac{d-1}{2}\right)}, \hspace{0.25cm}
	c_2 = -\frac{ \pi ^{\frac{d-1}{2}} }{60 (\nu+2)  \Gamma \left(\dfrac{d-1}{2}\right)}, \hspace{0.25cm} \cdots
\end{aligned}\end{equation}
where, $\d C = c_0 + c_1 (k\e)^2 + c_2 (k\e)^4 + \cdots$. These coefficients depend only on the choice of regulation scheme and not on the cut-off $\e$ at which the theory is regulated. Within such a scheme, with the regulated correlator, one needs to modify the effective Lagrangian appropriately to obtain the continuum power-law-obeying correlators. In conventional renormalization this is done by adding appropriate counter-terms in the Lagrangian. Following the general treatment of \cite{Pomoni:2008de}, we argue that in a large N theory the conformal invariance is broken by the running of double-trace couplings (which, as emphasized there, is a leading large N behaviour), unless the theory is at a conformal fixed point of all the double-trace couplings. Since we identify the alternative/standard quantizations with the UV/IR fixed points in the double-trace sectors, we are assured that no new counter-terms are generated for double-trace deformations. So, the corrections required in the regulated effective theory with certain UV cut-off can't be obtained by some double-trace counter terms. This argument is further strengthened by an explicit calculation with the inclusion of double-trace counter terms. As shown in various places in this paper, inclusion of any double-trace interaction in the Lagrangian (away from the fixed point values) necessarily modifies the correlators by addition of terms proportional to $k^{4\nu}, k^{6\nu}, \ldots$ -- which is not the same as the momentum space counter-terms that are present in the regulated theory. We believe that the inclusion of terms quadratic in the source, $J(k)$, of the operator, $\mO(k)$ in the Lagrangian provides the required correction that makes the correlators same as that of the continuum theory. Normally, in the partition function (which is computed with $J(k)=0$, as opposed to the generating function), one would think that such terms are inconsequential. However, such terms necessarily correct the generating function, $W[J] = \log Z[J]$, of the theory and hence all the correlators of the theory. Particularly, in the quadratic effective action that we have in the large N theory, we obtain the power-law 2-point functions with the inclusion of appropriate terms. Within Wilson-Polchinski fRG treatment, such terms are necessarily generated as we integrate out the degrees of freedom (\autoref{fig:jsq}).
\begin{figure}[t]
\begin{center}
\includegraphics[scale=0.7]{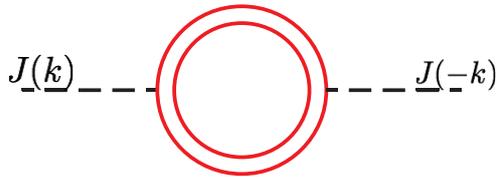}
\caption{Diagrams contributing to generation of terms quadratic in source, $J(k)$. As a standard convention throughout the paper, colored propagators denote `heavy' modes (see \autoref{sec:field-beta} for conventions used in Feynman diagrams).}
\end{center}
\label{fig:jsq}
\end{figure}

The bulk computation at finite radial cut-off, \eq{eq:exact-finite-partition-st}, automatically corresponds to the regulated field theory with the inclusion of such terms. However, we emphasise the need to differentiate the contribution of the regulation scheme from that of the quadratic $J$ term. In a regulated field theory with a double-trace deformation the regulation of the correlators (contact terms coming due to the regulation scheme) participates dynamically in the computation of the Feynman diagrams that gives rise to the rational fraction form of the correlator, \eq{dyson-schwinger-ft}, in the perturbed theory. The quadratic $J$ term corrects this correlator by an additive term (which cancels the regulation-scheme contact terms in absence of the perturbation). Analogously, in the bulk computation, we treat the two contributions separately. This is done by a deviation of the boundary wavefunctional, $\Psi$, from $C^*$ by some $\d C$ corresponding to the particular choice of scheme in the field theory. Then we use this wavefunctional in our Hubbard-Stratonovic transformation to describe the regulated, double-trace deformed field theory (as in \eqref{eq:dbl-trace-bulk-st-2} and \eqref{eq:dbl-trace-bulk-aq-2}). It is hence important to compute the $\b$-functions for the double-trace couplings using this prescription.

\section{Geometric interpretation: smeared Witten diagram}\label{sec:geo-interpret}
The above improvement of the AdS/CFT prescription at finite radial cut-off has a natural generalization in the limit of massive, $m R_{AdS} \gg 1$, bulk fields. It is known that in this limit, the field theory correlators are  approximated by geodesics between the points of operator insertions in the boundary, \cite{Balasubramanian:1999zv,Louko:2000tp} . 
Geodesic length between the points $(\e, x_1)$ and $(\e, x_2)$ in AdS is given by
\[
L_\e(x_1 - x_2)=\cosh^{-1}(1+ \f12 (|x_1 - x_2|/\e)^2)=  
2 \log[|x_1 - x_2|/\e] + 2 (\e/|x_1 - x_2|)^2 + O(\e/|x_1 - x_2|)^4
\]
This is related to the correlator $\lan\mO(x_1) \mO(x_2)\ran_\e$ for large $\Delta \approx m$ (with $R_{AdS}=1$) as ($\D$ is the operator dimension of $\mO$)
\be
G_\e(x_1- x_2)= \text{constant}~ \exp[- \Delta L_\e(x_1 - x_2)]
=   (1/|x_1 - x_2|)^{2\Delta}
e^{\left(1 + 2 \Delta (\e/|x_1 - x_2|)^2  +O(\e/|x_1 - x_2|)^4\right)}
\label{scaling-violation}
\ee
where the `constant'= $\e^{-2\Delta}$ (in accordance with
the dimension $[O(x)]= \Delta$, and Zamolodchikov's convention
$G(0,1)=1$). The corrections that appear in the exponential of the correlator above can be thought of as a regulation scheme for the correlator. It can be easily checked that this scheme obeys all the general discussion of \autoref{sub-sec:ft-corr-mom-sp} and has the momentum space counter-terms as discussed there.

Like the conventional GKPW prescription, this should also be understood as a limiting prescription which is well defined only in $\e\to0$ limit. Our finite radial cut-off modification to the GKPW prescription suggests that we need to modify the geodesic prescription too. Our source corresponding to the insertion of boundary operator $\mO$ at $x_1,x_2$ is $J(\vec{x}) = \d(\vec{x}-\vec{x}_1)+\d(\vec{x}-\vec{x}_2)$. Using the boundary condition, \eqref{eqn:jab-st} (with $\bfd=0$), we find that the bulk field, $\phi$, at finite radial cut-off in the momentum space is,
\begin{align}\label{field-momenta}
	\phi(k,\e_0) = \frac{2^{1-\nu } \e_0^{d/2} }{ \Gamma (\nu )} \left(e^{i \vec{k} \cdot \vec{x}_1}+e^{i \vec{k} \cdot \vec{x}_2} \right) k^{\nu } K_{\nu }(k \e_0)
\end{align}
where we have used $J(k) = \left(e^{i \vec{k} \cdot \vec{x}_1}+e^{i \vec{k} \cdot \vec{x}_2} \right)$. Similar to the law of superposition, we simply add the field due to the presence of one source at $\vec{x}=\vec{x}_1$ to that due to source at $\vec{x}=\vec{x}_2$. In position space, the field due to an individual source is given by,
\begin{align}
	\phi(k,\e_0) &= \frac{2^{1-\nu } \e_0^{d/2} }{ \Gamma (\nu )} e^{i \vec{k} \cdot \vec{x}_1} k^{\nu } K_{\nu }(k \e_0) \ \underrightarrow{\substack{\text{Fourier} \\ \text{transform}}} \nonumber \\
	& \frac{2^{d-1} \pi ^{\frac{d-2}{2}} \e_0^{-\frac{d}{2}-\nu }}{ (d+2 \nu -1)} \left( \frac{\Gamma \left(\frac{d}{2}\right) \Gamma \left(\frac{d}{2}+\nu \right)}{\Gamma \left(\frac{d+1}{2}\right) \Gamma (\nu )} \right) \Bigg( \left( 1 + \frac{\e_0^2}{\rho ^2}\right) \, _2F_1\left(\frac{d}{2},\frac{d}{2}+\nu ;-\frac{1}{2};-\frac{\rho ^2}{\e_0^2}\right) \nonumber \\
	& \hspace{5cm}-\left(2 (d+\nu )+ \frac{\e_0^2}{\rho ^2} \right) \, _2F_1\left(\frac{d}{2},\frac{d}{2}+\nu ;\frac{1}{2};-\frac{\rho ^2}{\e_0^2}\right) \Bigg)
\end{align}
This function is peaked around $\rho=0$, where $\vec{\rho} = \vec{x}-\vec{x}_1$, with a half-width of the order of $\e_0$.
\begin{figure}[H]
\begin{center}
\includegraphics[trim={1.2cm 0 0 0},clip]{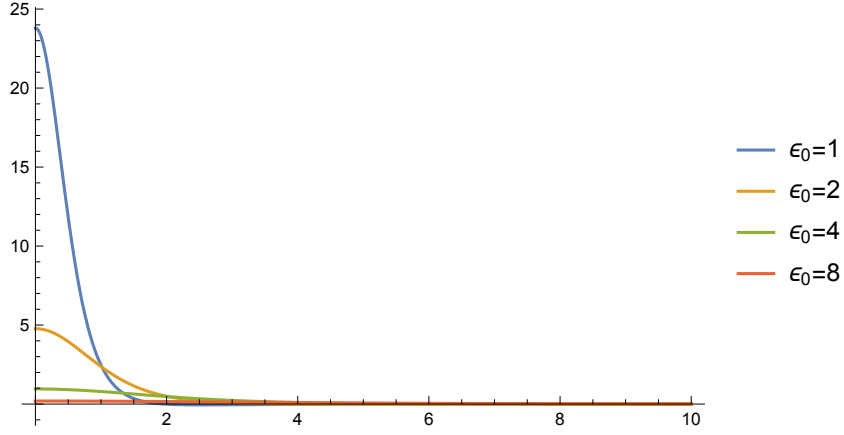}
\caption{Plots for the boundary fields at finite radial cut-off with double-centered delta function source.}
\end{center}
\label{fig:geodesic_regulated}
\end{figure}
This solution for $\phi_0$ corresponds to a distribution for $\phi_0$ smeared around $J(x)= \delta(\vec{x}- \vec{x}_1) + \delta(\vec{x}-\vec{x}_2)$. This is schematically represented by the right panel of the diagram, 
\autoref{fig:regulated-witten-diagram}. Note that since the correlator at any
cut-off surface is a pure power law by this device, {\large\sl the
  motion of the cut-off surface into the AdS bulk does not change the
  correlator}.

\begin{figure}[H]
\begin{center}
\includegraphics[scale=0.5]{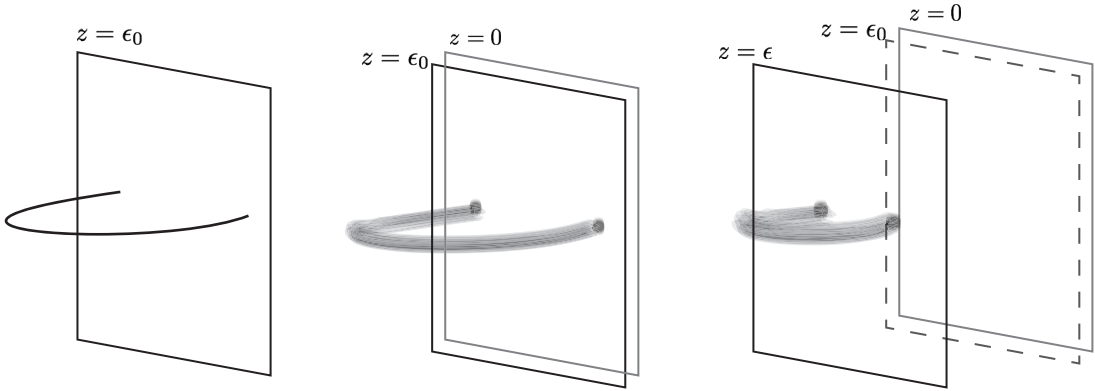}
\caption{(Left) Witten diagram for a delta-function boundary term corresponds to
a scaling violation, as in \eq{scaling-violation}. (Center) Witten
diagram with our smearing over the delta-function boundary condition
gives the pure power law. (Right) Smearing increases as one moves deeper in the radial direction. However, the exact correlator in both the centre and the right diagram are equal.}
\end{center}
\label{fig:regulated-witten-diagram}
\end{figure}

\section{Double trace perturbations\label{sec:pres}}
Having defined our fixed point theories with a finite cut-off and before we move on to computation of $\b$-function in dual bulk theory, we review (\cite{Witten:2001ua}) and extend the AdS/CFT dictionary for the derivative double-trace operators. We show that the same bulk field which is dual to a scalar primary operator $\mO$ of scaling dimension $\D$ also describes the physics of derivative multi-trace operators with an appropriately modified boundary condition that we discuss in this section.
    
Our action with a double-trace perturbation and inclusion of a source term is given by \eqref{eq:perturbation-Langrangian-FT}
    \begin{align}
	S = S_0 + \half \int d^dk \  \mO \bfd(\del^2) \mO(x)  - \int d^dx \ J(x) \mO(x) \nonumber
    \end{align}
Since the bulk computations will give us different $\b$-functions, to differentiate between the two sets of couplings we have denoted the couplings used in the bulk calculations by $\bfd$ instead of $\fd$ for the dimensionful couplings, and $\bfnd$ instead $\fnd$ for the dimensionless couplings. We are using the same notation for $\bfd(\del^2)$ as in \eqref{eq:packaged-coupling-FT}. In the subsequent discussions, we work in momentum space,
\[
    \bfd(k^2) \ = \ \bfd_0 + \bfd_2 k^2 + \bfd_4 k^4 + \ldots
\]

    We use Hubbard-Stratonovich trick to write the perturbation terms above as,
    \begin{align}\label{eq:Hub-Strat}
	\exp\left[\int J(k) \mO(-k)\right. -& \left. \int \frac{\bfd(k^2)}{2} \mO(k) \mO(-k) \right] \nn\\
	& = \int \mathcal{D}\tilde{\phi}\,\exp \left[ \int \dfrac{\left(\tilde{\phi}-J\right)_{k} \left(\tilde{\phi} - J \right)_{-k}}{2 \bfd(k^2)} 
	+ \int \tilde{\phi}(k) \mO(-k) \right]
    \end{align}

\paragraph{Standard Quantization}
Using \eqref{eq:Hub-Strat}, and the statement of duality for standard quantization at finite radial cut-off given by the wavefunctional \eqref{eq:wavefunctional-correct-st}, we obtain a bulk partition function dual to the double-trace perturbed field theory,
    \begin{align}\label{eq:dbl-trace-bulk-st}
	Z_+[J,\bfd(k^2)] \ = \ \int \mathcal{D}\phi \ \exp &\left[ - S_b  - \int \limits_{z=\e_0} d^dk \ \sqrt{\g_0} \ \frac{\left( \phi + \afxd \cdot B_{ST}^*(k\e_0) \ \e_0^{d-\D_+} J \right)_{k}^2 }{2 \afxd(k\e_0) \left( 1 - {B^*_{ST}}^2 \ \afxd \dfrac{\bfd(k^2)}{\e_0^{2\nu}}\right)}  \right. \nonumber \\
            &\left.- \int d^dk \frac{\sqrt{\g_0}}{2} \phi_k \DCT(\e_0 k) \phi_{-k} \right]
    \end{align}
        Variational principle imposes following condition at the boundary $z=\e_0$,
    \begin{align}
	\pi(k,\e_0) - \sqrt{\g_0} \ \frac{ \left( \phi + \afxd \cdot B_{ST}^*(k\e_0) \ \e_0^{d-\D_+} J \right)_{k}}{ \afxd(k\e_0) \left( 1 - {B^*_{ST}}^2 \ \afxd \dfrac{\bfd(k^2)}{\e_0^{2\nu}}\right)} - \sqrt{\g_0} \ \DCT(\e_0 k) \phi(k,\e_0) = 0
\label{variational}
    \end{align}
    where, $\pi(k,z) = \sqrt{g} \ \del^z\phi(k,z)$ is the conjugate momentum of the bulk field.\\
    Using the near boundary expansion of the bulk field $\phi(k,z)$,
    \begin{equation}\label{eq:sol-bulk-wave}
	\phi(k,z) = z^{d-\D_+} \ a(k) \left( 1 - \frac{(kz)^2}{2^2 (\nu-1)} + \cdots \right) + z^{\D_+} \ b(k) \left( 1 + \frac{(kz)^2}{2^2 (\nu+1)} + \cdots \right)
    \end{equation}
    the boundary condition becomes,
    \begin{align}\label{eqn:jab-st}
	J(k) = 2 \nu \bfd(k^2) b(k) + \ a(k) \ \footnotemark
    \end{align}
    \footnotetext{This equation is correct to all orders with the inclusion of all the correct counterterms that we have derived at finite cut-off, viz., the values of $B^*_{ST}, \afxd, \DCT$.}%
    In the above expression, in the $\e_0\to0$ limit, $b(k)$ is the expectation value of the operator $\mO$, and $a(k)$ is the source. The above expression can be rewritten as,
    \begin{align}\label{eq:gen-witten-st}
	& a(k) \ = \ J(k) - 2 \nu \bfd(k^2) \ b(k) \ = \ J(k) - 2 \nu \bfd(k^2) \langle \mO(k) \rangle \nonumber \\
	\equiv \ & a(x) \ = \ J(x) - 2 \nu \Big( \bfd_0 \langle \mO(x) \rangle + \bfd_2 \langle \del^2 \mO(x) \rangle + \bfd_4 \langle \del^4 \mO(x) \rangle + \ldots \Big)
    \end{align}
    
    IR boundary condition in the bulk at $z=\infty$ imposes an additional condition on the on-shell field $\phi(k,z)$. In the pure AdS geometry, demanding the regularity of the field at IR determines $b(k)$ in terms of $a(k)$,
    \[
    	b(k) = 2^{-2\nu} k^{2\nu} \frac{\Gamma(-\nu)}{\Gamma(\nu)} \cdot a(k)
    \]
    So the improved relationship between the boundary value of the bulk field, $\phi(k,\e_0)$, and the field theory source for the dual operator $\mO$, in the absence of the double-trace deformation, $\bfd(k^2)$, is
    \begin{align}\label{source-field}
    	\phi(k,\e_0) = \e_0^{d-\D_+} J(k) \left[ \left( 1 - \frac{(k\e_0)^2}{2^2 (\nu-1)} + \cdots \right) + \left(\frac{k\e_0}{2}\right)^{2\nu} \frac{\Gamma(-\nu)}{\Gamma(\nu)} \left( 1 + \frac{(k\e_0)^2}{2^2 (\nu+1)} + \cdots \right) \right]
    \end{align}
    In the limit, $\e_0\to0$, this gives back the well known GKPW prescription between the field and the source, $\lim\limits_{\e_0\to0} \e_0^{\D_+ - d} \phi(k,\e_0) = J(k)$. This is a reaffirmation of the limiting $\d$-function prescription, \eqref{st-GKPW}, originally known in the correspondence.\\
    In the presence of the double-trace deformation this relation gets modified to,
    \begin{align}\label{source-field-double-trace}
    	\phi(k,\e_0) = \e_0^{d-\D_+} J(k) \frac{ \left[ \left( 1 - \frac{(k\e_0)^2}{2^2 (\nu-1)} + \cdots \right) + \left(\frac{k\e_0}{2}\right)^{2\nu} \frac{\Gamma(-\nu)}{\Gamma(\nu)} \left( 1 + \frac{(k\e_0)^2}{2^2 (\nu+1)} + \cdots \right) \right] }{ 1 + 2^{1-2\nu} \ \bfnd(k^2 \e_0^2) \ \left(k\e_0\right)^{2\nu} \ \frac{\nu \Gamma(-\nu)}{\Gamma(\nu)} }
    \end{align}
    
\paragraph{With the regulator counter-terms} Since we are particularly interested in field theories that are regulated at short distances in position space (or equivalently, have certain counter-terms in the momentum space) it is also important that we establish our duality for the double-trace perturbations with the inclusion of such regulators, \eq{eq:Sextra-st}.
       \begin{align}\label{eq:dbl-trace-bulk-st-2}
	Z_+[J,\bfd(k^2)] \ = \ \int \mathcal{D}\phi \ \exp &\left[ - S_b  - \half \int \limits_{z=\e_0} d^dk \ \sqrt{\g_0} \ \left( \frac{1 - \d C \cdot \dfrac{\bfd(k^2)}{\e_0^{2\nu}}}{\afxd \left( 1 - \left(\d C + {B^*_{ST}}^2 \ \afxd \right) \dfrac{\bfd(k^2)}{\e_0^{2\nu}} \right)} \phi_k \phi_{-k}  \right. \right. \nonumber \\[5pt]
	& \hspace{-3.5cm}\left.+ 2 \dfrac{ B^*_{ST} \ \e_0^{d-\D_+}}{\left( 1 - \left(\d C + {B^*_{ST}}^2 \ \afxd \right) \dfrac{\bfd(k^2)}{\e_0^{2\nu}} \right)} J_k \phi_{-k} + \dfrac{\left( \d C + \afxd \cdot {B^*_{ST}}^2 \right) \e_0^{2(d-\D_+)}}{ \left( 1 - \left(\d C + {B^*_{ST}}^2 \ \afxd \right) \dfrac{\bfd(k^2)}{\e_0^{2\nu}} \right)} \ J_k J_{-k} \right) \nonumber\\[7pt]
           &\left. \hspace{2cm} - \int \limits_{z=\e_0} d^dk \frac{\sqrt{\g_0}}{2} \phi_k \DCT(\e_0 k) \phi_{-k} \right]
    \end{align}
    Variational principle imposes following condition at the boundary $z=\e_0$,
    \begin{align}
	    \pi(k,\e_0) - \frac{\sqrt{\g_0}}{\afxd} \left( \frac{1 - \d C \cdot \dfrac{\bfd(k^2)}{\e_0^{2\nu}}}{1 - \left(\d C + {B^*_{ST}}^2 \ \afxd \right) \dfrac{\bfd(k^2)}{\e_0^{2\nu}}} \right) \phi_k &- \sqrt{\g_0}\left( \frac{\e_0^{d-\D_+} B^*_{ST}}{ 1 - \left(\d C + {B^*_{ST}}^2 \ \afxd \right) \dfrac{\bfd(k^2)}{\e_0^{2\nu}}} \right)J_{k} \nonumber\\
	&- \sqrt{\g_0} \ \DCT(\e_0 k) \phi(k,\e_0) = 0
    \end{align}

    the boundary condition becomes,
    \begin{align}\label{eqn:jab-st-2}
		J(k) = 2 \nu \bfd(k^2) b(k) + \left(1- \frac{\bfd(k^2)}{\e_0^{2\nu}} \d C(k\e_0) \right) \ a(k) 
    \end{align}
    
In the double-trace perturbed theory the exact two point function, $\langle \mO(k) \mO(-k) \ran_{\bfd}$ is given by the summing over all the connected diagrams. Since the bulk partition function of the perturbed theory, \eq{eq:dbl-trace-bulk-st} or \eq{eq:dbl-trace-bulk-st-2}, is quadratic in bulk fields $\phi_k$, we can perform the gaussian integral exactly and compute the 2-point function from the resulting generating function,
\begin{align}\label{eq:corr-def-st}
	\langle \mO(k) \mO(-k) \ran^{(+)}_{\bfd} = \dfrac{G^{\e_0}_{(+)}(k)}{1+\bfd(k^2) G^{\e_0}_{(+)}(k)}
\end{align}
for any value of the coupling $\bfd(k^2)$. Here $G^{(\e_0)}_+$ is given by either \eqref{eq:exact-2-pt-finite-st} or \eqref{eq:counter-2pt-st}.\footnote{Note that we have dropped the contribution coming from the quadratic $J$ explained in \autoref{sub-sec:choice-of-regulation-bulk} as we won't need them for the $\b$-function calculation, but we should remember their presence.}

\paragraph{Alternative Quantization}
From the duality for alternative quantization without double-trace perturbation \eqref{eq:wavefunctional-correct-aq} and \eqref{eq:Hub-Strat}, the bulk dual to double-trace deformed alternative quantized theory is,
    \begin{align}\label{eq:dbl-trace-bulk-aq}
	Z_{-}[J,\bfd(k^2)] &= \int \mD \P \exp \left( -S^{(-)}_0 + \int d^dk \ J(k) \mO(-k) - \int d^dk \ \frac{\bfd(k^2)}{2} \mO(k) \mO(-k) \right) \nonumber \\
	&= \int\limits_{z\ge\e_0} \mD \phi \exp \left[ -S_b  - \int\limits_{z=\e_0} d^dk \frac{\sqrt{\g_0}}{2} \left( \frac{ {B_{AQ}^*}^2 \ \bfd(k^2) \e_0^{2\nu} }{ 1 - C_{AQ}^*\ \bfd(k^2) \e_0^{2\nu} } + \DCT(\e_0k) \right) \phi_k \phi_{-k}   \right. \nonumber \\
	& \hspace{3cm} - \int\limits_{z=\e_0} d^dk \sqrt{\g_0} \left( \frac{ B_{AQ}^* }{ 1 - C_{AQ}^*\ \bfd(k^2) \e_0^{2\nu} } \right) \e_0^{d-\D_-} \phi_k J_{-k} \nonumber \\
          & \hspace{4cm} \left. - \half \int\limits_{z=\e_0} d^dk \sqrt{\g_0} \left( \frac{ {C_{AQ}^*} }{ 1 - C_{AQ}^*\ \bfd(k^2) \e_0^{2\nu} }\right) \e_0^{2(d-\D_-)} J_k J_{-k}\right] 
    \end{align}
        Variation of the fields on the boundary $z=\e_0$ imposes the condition,
    \[
	\pi(k,\e_0) - \sqrt{\g_0} \ \phi(k,\e_0) \left( \frac{ {B_{AQ}^*}^2 \ \bfd(k^2) \e_0^{2\nu} }{ 1 - C_{AQ}^*\ \bfd(k^2) \e_0^{2\nu} } + \DCT(\e_0k) \right) = \sqrt{\g_0}\ \left( \frac{ B_{AQ}^* }{ 1 - C_{AQ}^*\ \bfd(k^2) \e_0^{2\nu} } \right) \e_0^{d-\D_-} J(k)
    \]
    Using the near boundary expansion of the bulk field $\phi(k,z)$ in the boundary condition we get,
    \footnote{$\phi(k,z) = z^{d-\D_-} \ a(k) \left( 1 + \frac{(kz)^2}{2^2 (\nu+1)} + \cdots \right) + z^{\D_-} \ b(k) \left( 1 - \frac{(kz)^2}{2^2 (\nu-1)} + \cdots \right)$ where $a(k)$ is the coefficient of normalizable part and hence the source for alternative quantization. Also the expression in \eqref{eq:jab-aq} is exact to all orders.}%
    \begin{align}\label{eq:jab-aq}
	J(k) = 2 \nu \ a(k) - \bfd(k^2) b(k)
    \end{align}
    which can be rewritten as,
    \begin{align}\label{eq:gen-witten-aq}
	a(k) & =  \frac{1}{2\nu} \left( J(k) + \bfd(k^2) b(k) \right) = \frac{1}{2\nu} \left( J(k) + \bfd(k^2) \langle \mO(k) \rangle \right) \nonumber \\
	\equiv \ a(x)  &= \frac{1}{2\nu} \left( J(x) + \bfd_0 \langle \mO(x) \rangle + \bfd_2 \langle \del^2 \mO(x) \rangle + \bfd_4 \langle \del^4 \mO(x) \rangle + \ldots \right)
    \end{align}
    
    As in the standard quantization, demanding regular IR boundary condition in the pure AdS bulk geometry, at $z=\infty$, determines $b(k)$ in terms of $a(k)$,
    \[
    	b(k) = 2^{2\nu} k^{-2\nu} \frac{\Gamma(\nu)}{\Gamma(-\nu)} \cdot a(k)
    \]
    So the improved relationship between the boundary value of the bulk field, $\phi(k,\e_0)$, and the field theory source for the dual operator $\mO$, now of dimension $\D_-$, is 
    \begin{align}\label{source-field-a}
    	\phi(k,\e_0) = \e_0^{\D_-} J(k) \left(\frac{k}{2}\right)^{-2\nu} \frac{ \left[ \left(\frac{k\e_0}{2}\right)^{2\nu} \left( 1 + \frac{(k\e_0)^2}{2^2 (\nu+1)} + \cdots \right) +  \frac{\Gamma(\nu)}{\Gamma(-\nu)} \left( 1 - \frac{(k\e_0)^2}{2^2 (\nu-1)} + \cdots \right) \right] }{ 2\nu - 2^{2\nu} \ \bfnd(k^2 \e_0^2) \ \left(k\e_0\right)^{-2\nu} \ \frac{\Gamma(\nu)}{\Gamma(-\nu)} }
    \end{align}
    which, again limits to the known relationship between the source and the normalizable part of the bulk field, $J(k) = 2\nu a(k)$ in the $\e_0\to0$ limit in the absence of the double-trace deformations.
    
    \paragraph{With the regulator counter-terms} If we however start with \eq{eq:Sextra-aq}, then,
    \begin{align}\label{eq:dbl-trace-bulk-aq-2}
	Z_{-}[J,\bfd(k^2)] &= \int \mD \P \exp \left( -S^{(-)}_0 + \int d^dk \ J(k) \mO(-k) - \int d^dk \ \frac{\bfd(k^2)}{2} \mO(k) \mO(-k) \right) \nonumber \\
	&= \int\limits_{z\ge\e_0} \mD \phi \exp \left[ -S_b  - \half \int\limits_{z=\e_0} d^dk \sqrt{\g_0} \left( \frac{{B^*_{AQ}}^2 \ \e_0^{2\nu} \; \bfd(k^2)}{1- \e_0^{2\nu} \; \bfd(k^2) (C^*_{AQ}+\d C) } + \DCT(\e_0k) \right) \phi_k \phi_{-k}   \right. \nonumber \\
	& \hspace{3cm} - \int\limits_{z=\e_0} d^dk \sqrt{\g_0} \ \frac{B_{AQ}^*}{1- \e_0^{2\nu} \; \bfd(k^2) (C^*_{AQ}+\d C)} \e_0^{d-\D_-} \phi_k J_{-k} \nonumber \\
	& \hspace{3cm} \left.- \half \int\limits_{z=\e_0} d^dk \sqrt{\g_0} \ \frac{C_{AQ}^* + \d C}{1- \e_0^{2\nu} \; \bfd(k^2) (C^*_{AQ}+\d C)} \ \e_0^{2(d-\D_-)} J_k J_{-k}\right] 
    \end{align}
	which leads to boundary condition,
    \begin{align}
	\pi(k,\e_0) - \sqrt{\g_0} \ \phi(k,\e_0) \left( \frac{{B^*_{AQ}}^2 \ \e_0^{2\nu} \; \bfd(k^2)}{1- \e_0^{2\nu} \; \bfd(k^2) (C^*_{AQ}+\d C) } + \DCT(\e_0k) \right) =\nonumber \\
	 \sqrt{\g_0}\ \frac{B_{AQ}^*}{1- \e_0^{2\nu} \; \bfd(k^2) (C^*_{AQ}+\d C)} \ \e_0^{d-\D_-} J(k)
    \end{align}    
    \begin{align}\label{eq:jab-aq-2}
	J(k) = 2 \nu \left(1 - \e_0^{2\nu} \bfd(k^2)\ \d C (k\e_0) \right) \ a(k) - \bfd(k^2) b(k)
    \end{align}
    
As in standard quantization, the 2-point function is evaluated exactly by integrating out \eqref{eq:dbl-trace-bulk-aq} or \eq{eq:dbl-trace-bulk-aq-2},
\begin{align}\label{eq:corr-def-aq}
	\langle \mO(k) \mO(-k) \ran^{(-)}_{\bfd} = \dfrac{G^{\e_0}_{(-)}(k)}{1+\bfd(k^2) G^{\e_0}_{(-)}(k)}
\end{align}\\
    Equations \eqref{eq:gen-witten-st}, \eq{eqn:jab-st-2}, \eqref{eq:gen-witten-aq} and \eqref{eq:jab-aq-2}  are our proposed generalisation of the boundary prescription originally given by \cite{Witten:2001ua} for the derivative multi-trace deformations around a conformal field theory in standard and alternative quantization, respectively. These have the same structure as we had found for the field theory correlators in \autoref{sub-sec:ft-corr-mom-sp}.\\
    For even more general higher-derivative multi-trace operators, we expect that the above formulae generalises as long as we include all the derivative terms inside the expectation values. Corresponding computation for triple-trace operators without derivatives is done in \cite{Aharony:2015afa}, and we think the generalisation shouldn't be difficult.
    
\section[\texorpdfstring{Holographic computation of $\b$-functions}{Holographic computation of beta-functions}]{\label{sec:hol-RG}Holographic computation of $\b$-functions}

Having established the duality for the double-trace operators in
previous section, we know that the couplings of the field theory
double-trace operators are contained in the coefficient of the $\phi_k
\phi_{-k}$ in the boundary part of the bulk action
\eqref{eq:dbl-trace-bulk-st},\eqref{eq:dbl-trace-bulk-aq}. AdS/CFT naturally
incorporates a holographic version of RG flow, because of the
correspondence between the radial coordinate in the bulk and the
energy scale in the boundary field theory, see, e.g.,
\cite{deBoer:1999tgo, Verlinde:1999xm,
  Balasubramanian:1999jd, deBoer:2000cz, Akhmedov:2002gq}. Holographic
Wilsonian RG flow of double-trace operators without derivatives was
considered in \cite{Heemskerk:2010hk, Faulkner:2010jy}, which was
generalised in \cite{Elander:2011vh} to double trace operators with
derivatives. In the following we essentially build up on the treatment
in \cite{Elander:2011vh}. For other relevant work on renormalization
of multi-trace operators from holographic and field theoretic
viewpoints, see, e.g.  \cite{Pomoni:2008de, Laia:2011wf,
  Grozdanov:2011aa, Balasubramanian:2012hb, Aharony:2015afa}).

An essential feature of the AdS/CFT correspondence is the connection between the energy scale of the conformal field theory (CFT) and the
radial coordinate of the AdS dual. More precisely, AdS/CFT states that the bulk partition function in Euclidean AdS, defined with a radial
cut-off $r=r_0$, equals the dual field theory partition function with a UV momentum cut-off $\Lambda$ given in terms of $r_0$ (for large
$\Lambda$, $\Lambda=r_0/R_{AdS}^2$ \cite{Susskind:1998dq}). A corollary of this statement, in the semi-classical limit, is that the
running of field theory couplings is identified with the radial dependence of classical field configurations in the dual gravitational theory (see, e.g., \cite{deBoer:1999tgo,  Verlinde:1999xm, Balasubramanian:1999jd, deBoer:2000cz}). Motivated by this feature, in \cite{Heemskerk:2010hk, Faulkner:2010jy}, the near-boundary degrees of freedom in the bulk are identified with the \emph{heavy}/short-distance modes  of the dual field theory. They work in \emph{probe approximation} with a fluctuating field $\phi(x,z)$ on a fixed AdS background given by,
\begin{equation}\label{eq:AdS-metric}
  ds^2 = \frac{1}{z^2} \left( dz^2 + \eta_{\mu \nu} dx^\mu dx^\nu \right)
\end{equation}
Integration of the near boundary modes in the bulk gives a new holographic version of Wilsonian effective action in the field theory. Stated mathematically,
\begin{align}\label{eqn:Zbulk}
    && Z_{bulk,\e_0} &= \int_{z\ge\e_0} \mathcal{D} [\phi] e^{-\mathcal{S}[\phi]} \nonumber \\
    && & = \int \mathcal{D}\phi|_{z>\e}\ \mathcal{D}\tilde{\phi}\ \mathcal{D}\phi|_{\e_0\le z<\e}\quad e^{-\mathcal{S}[\phi]|_{z>\e}} e^{-\mathcal{S}[\phi]|_{z<\e}} \nonumber \\
    && &= \int \mathcal{D}\tilde{\phi} \ Z_{bulk,\e}(\e,\tilde{\phi}) Z_{UV}(\e,\tilde{\phi})
\end{align}
The role of $Z_{UV}$ is an addition of a boundary wavefunctional, $\Psi[\phi_0;\e_0]$ to the bulk action at the new cutoff $z=\e$, $Z_{bulk,\e}$. This, in the AdS/CFT dictionary has the interpretation of addition of higher-trace terms in the field theory, as discussed in \autoref{sec:pres}. Following Wilsonian principles, same as in the field theory computations, we demand,
\begin{align}\label{eq:Bulk-Wilson}
	&& \dfrac{d}{d\e}Z_{bulk,\e_0} &= 0 \nonumber \\
	&& \Rightarrow \int \mathcal{D} \tilde{\phi} \left( \partiald{Z_{bulk,\e}}{\e} \ Z_{UV} \right. &+ \left. Z_{bulk,\e} \ \partiald{Z_{UV}}{\e} \right) = 0
\end{align}
here, the evolution of $Z_{UV}$ can be computed using the Hamiltonian corresponding to radial slicing,
    \begin{align}\label{eqn:evolution}
	\partiald{Z_{UV}}{\e}(\tilde{\phi},\e) &= -H(\tilde{\phi},\tilde{\pi}) Z_{UV}(\e,\tilde{\phi})
    \end{align}
which we will refer to as radial Schr\"{o}dinger evolution equations. Here $\tilde{\pi}= -i \kappa^2 \ \delta/\delta \tilde{\phi}$. In general, $Z_{UV}$ contains the details of the various field theory couplings which enables us to compute the $\b$-functions of these couplings using \eqref{eqn:evolution}. These ideas have been worked out for the bulk duals of double-traced deformed field theories \eqref{eq:dbl-trace-bulk-st}, \eqref{eq:dbl-trace-bulk-aq} in \autoref{app:Eff-Act}. We only quote the final $\b$-functions here,
\paragraph{Standard Quantization:} 

Working with the bulk action, \eqref{eq:dbl-trace-bulk-st-2}, which is dual to the regulated field theory and keeping in mind the subtleties that we remarked upon in the \autoref{sub-sec:choice-of-regulation-bulk}, we get the $\b$-function equation,
    \begin{align}\label{eq:st-bulk-beta-dC}
	\e \del_\e \bfnd =&\ \bfnd^2 \times \Bigg( {B^*_{ST}}^2 \ {\afxd}{}^2 \left( k^2 \e^2+ m^2 + \DCT (d-\DCT)-\e \del_\e\DCT \right) - \e \del_\e \d C  \nonumber \\
	& \hspace{0.35cm} + \frac{(\d C)^2}{{B^*_{ST}}^2 \afxd{}^2}  \left( \e \del_\e \bfnd^*-1 + \afxd (d-2 \DCT)+\afxd{}^2 \left( k^2 \e^2 + m^2 + \DCT (d-\DCT)-\e \del_\e \DCT \right) \right) \nonumber \\
	& \hspace{-0.25cm} - \d C \left( -\frac{2 \e \del_\e B^*_{ST}}{B^*_{ST}} - \frac{2 \ \e \del_\e{\afxd}}{\afxd} - 2 {\afxd} \left( k^2 \e^2 + m^2 + \DCT (d-\DCT)-\e \del_\e \DCT \right) -d+2 \DCT \right) \Bigg) \nonumber \\[7pt]
	    & +\hspace{-1.8pt}  \bfnd \Bigg(-\frac{2 \ \e \del_\e{B^*_{ST}}}{{B^*_{ST}}} - \frac{2 \ \e \del_\e{\afxd}}{\afxd} -2 {\afxd} \left( k^2 \e^2+ m^2 + \DCT (d-\DCT)-\e \del_\e\DCT\right) -d+2 \DCT(k \e) \nonumber \\
	& \hspace{-0.25cm}  - 2 \dfrac{\d C}{{B^*_{ST}}^2 \afxd{}^2} \left( \e \del_\e \bfnd^*-1 + \afxd (d-2 \DCT)+\afxd{}^2 \left( k^2 \e^2 + m^2 + \DCT (d-\DCT)-\e \del_\e \DCT \right) \right) \Bigg) \nonumber \\[7pt]
	&\hspace{1.5cm} + \ \frac{ \e \del_\e{\afxd} + \afxd \ (d-2 \DCT)+\afxd{}^2 \left( k^2 \e^2+ m^2 + \DCT (d-\DCT)-\e \del_\e\DCT\right) -1}{{B^*_{ST}}^2 \ {\afxd}{}^2}
    \end{align}
\paragraph{Alternative Quantization:} Bulk action, \eq{eq:dbl-trace-bulk-aq-2}, corresponds to the regulated theory,
	\begin{align}\label{eq:aq-bulk-beta-dC}
		\e \del_\e \bfnd =& \frac{1}{{B^*_{AQ}}^2} \Bigg[\bfnd^2 \bigg(2 {B^*_{AQ}} \ \e \del_\e {B^*_{AQ}} \left( {C^*_{AQ}} + \d C \right) -{B^*_{AQ}}^2 \left(\e \del_\e {C^*_{AQ}} + \e \del_\e \d C+ \left({C^*_{AQ}} + \d C \right) (d-2 \DCT)\right) \nonumber \\
		&\hspace{4cm}  - {B^*_{AQ}}^4 + \left( {C^*_{AQ}} + \d C \right)^2 \left(\DCT (d-\DCT)-\e \del_\e \DCT+k^2 \e^2+m^2\right) \bigg) \nonumber \\[5pt]
		&\hspace{-0.85cm}+\bfnd \left(-2 {B^*_{AQ}} \ \e \del_\e {B^*_{AQ}} +{B^*_{AQ}}^2 (d-2 \DCT)-2 \left( {C^*_{AQ}} + \d C \right) \left(\DCT (d-\DCT)-\e \del_\e \DCT+k^2 \e^2+m^2\right)\right) \nonumber \\[5pt]
		&\hspace{3cm} + \DCT (d-\DCT)-\e \del_\e \DCT+k^2 \e^2+m^2 \Bigg]
	\end{align}

We have listed the $\b$-function equations for individual couplings $\bfnd_i$ in \eqref{eq:st-bulk-indiv-beta-2} and \eqref{eq:aq-bulk-indiv-beta-2}. One can note that they follow the same general structure as the $\b$-functions computed from the field theory. Although, even for the same choice of the regulator (or equivalently, $\d C$) at a given cut-off, the $\b$-functions are different. We associate this additional `scheme-dependence' of the $\b$-functions with reparametrization in the space of couplings as explained in \autoref{sec:rational-fraction}.


\section[\texorpdfstring{$\b$-function for double-trace operators from field theory}{Beta-function for double-trace operators from field theory}]{$\b$-function for double-trace operators from field theory}
\label{sec:field-beta}
  
\subsection[\texorpdfstring{Warming up: $\b$-function of $\fd_0$}{Warming up: Beta-function of f0}]{Warming up: $\b$-function of $\fd_0$}

Before we get into a full-fledged calculations of $\b$-function for
general double trace couplings mentioned above, let us first describe,
following \cite{Witten:2001ua, Vecchi:2010jz}, the Wilsonian
computation of the $\b$-function in the space of the single coupling
$\fd_0$. Double-trace perturbations without derivatives,
i.e. \eq{all-double-trace} with only $\fd_0 \ne 0$ and their
renormalizations have been discussed extensively in the literature
see, e.g.  \cite{Vecchi:2010jz, Witten:2001ua, Heemskerk:2010hk, Faulkner:2010jy, Elander:2011vh, Laia:2011wf, Vecchi:2010dd, Pomoni:2008de, Grozdanov:2011aa, Balasubramanian:2012hb,
  Aharony:2015afa, Akhmedov:2002gq}. 

Let us consider a double-trace perturbation given by,
\begin{align}
S = S_{CFT} + \frac{f_0}{2} \int d^dx \ \mO^2(x) 
\label{non-deriv}
\end{align}
 The single-trace operator $\mO(x)$ is a primary  of conformal dimension $\D_- = \dfrac{d}{2}-\nu$ at the fixed point given by $\fd_i=0$. The double-trace operator will then be a relevant operator with dimension (at leading large $N$). \footnote{This makes the theory at
  $f_0=0$ a UV CFT. In later sections discussing the holographic setup, we will identify this CFT with the so-called `alternative quantization'. However, we keep our subsequent analysis more general and won't use any specific value of $\D$. Only in \eqref{eq:beta-dimensionless-f0-only} do we use the specific value in \eq{relevant}}
\begin{align}
\Delta_{\mO^2}= 2 \D \equiv d- \nu, \; \nu>0
\label{relevant}
\end{align} 
In \cite{Witten:2001ua} $\b$-function for $f_0$ was computed for a marginal double-trace deformation. This was generalised in \cite{Vecchi:2010jz} to arbitrary $\Delta_{\mO^2}$, where a Wilsonian RG using real space integration shells was used. See also \cite{vanRees:2011fr}, and \cite{Aharony:2015afa} for a general perspective. Partition function of the deformed theory is given by,
\begin{align}\label{ft-partition}
Z= \int \mD \Phi \; e^{-S[\Phi]} = \int \mD \Phi \; e^{-S_{CFT}[\Phi]} \left( 1 - \frac{\fd_0}{2} \int d^dx \ \mO^2(x) + \frac{\fd_0^2}{4\cdot2!} \int d^dx\: d^dy \; \mO^2(x) \mO^2(y) - \ldots \right)
\end{align}
Here, $\Phi$ are the `fundamental fields' in the theory. 
The omitted terms in \eq{ft-partition} organise in themselves in form of a Dyson-Schwinger sum in the final answer.
If we regulate the theory at some cut-off $a$, such that the correlator $\lan \mO(x) \mO(y) \ran$ vanishes for $|x-y| \le a$, we can write (for more general treatment see \eq{regulator} and the discussion in Section \ref{details-FT}) 
\begin{align}\label{vecchi-regulator}
G_a(w) =\lan \mO(x) \mO(x+w) \ran_a = \frac{\Theta(|w|/a-1)}{|w|^{2 \D}} 
\end{align}
this regulator is also used in \cite{Vecchi:2010jz} (see Section \ref{details-FT}, especially \eq{eq:Regulated-Theta} for other choices). As explained in detail in following subsection (\autoref{fig:pert_heavy}), we can rewrite the third term in parenthesis in \eq{ft-partition} as,
\begin{align}\label{eq:G-exp}
  \frac{\fd_0^2}{4\cdot2!} \int d^dx\: d^dw \; \mO^2(x) \mO^2(x+w) &= \frac{\fd_0^2}{2!}  \int d^dx\: d^dw \; \mO(x) \; G_a(w) \; \mO(x+w) \nonumber \\
  &= \frac{\fd_0^2}{2!}  \int d^dx\: d^dw \;  \mO(x) \; \Big( G_{a'}(w) + (a-a') G'_{a'}(w) \nonumber \\
  & \hspace{2cm} +  \frac{(a-a')^2}{2} G''_{a'}(w) + \cdots \Big) \; \mO(x+w)
\end{align}
In \eqref{eq:G-exp} we have omitted the terms that are suppressed in the large $N$ limit (see \autoref{subsec:exactness}).
In the simple case of $\Theta$-function cut-off as in \eq{vecchi-regulator}, it can be written more simply as,
\begin{align}\label{eq:Th-exp}
   \frac{\fd_0^2}{4\cdot2!} \int_a d^dx\: d^dw \; \mO^2(x) \mO^2(x+w) &= \frac{\fd_0^2}{4\cdot2!} \Bigg( \int_{a'} d^dx\: d^dw \; \mO^2(x) \mO^2(x+w) \nonumber \\
   & \hspace{1.5cm}+ 4 \int^{a'}_a d^dx\: d^dw \; \mO^2(x) \mO^2(x+w) \Bigg) \nonumber \\
   &= \frac{\fd_0^2}{4\cdot2!} \Bigg( \int_{a'} d^dx\: d^dw \; \mO^2(x) \mO^2(x+w) \nonumber \\
   & \hspace{1.5cm}+ 4 \int^{a'}_a d^dx\: d^dw \; \mO(x) \:  \dfrac{1}{|w|^{2\D}} \: \mO(x+w) \Bigg)
\end{align}
The factors of 4 in both \eqref{eq:G-exp} and \eqref{eq:Th-exp} are due to 4 possible combinations of contractions between $\mO(x)$ and $\mO(x+y)$. While the first term in \eqref{eq:Th-exp} is the standard contribution for a new theory defined at cut-off $a'$, the second term corrects the value of $\fd_0$ in \eqref{ft-partition}. In second term on RHS of \eq{eq:Th-exp}, expanding $\mO(x+w)$ in a Taylor series
\begin{align}\label{taylor}
 \frac{\fd_0^2}{2} \int d^dx\;\left( \mO^2(x) \int_a^{a'} d^dw \frac{1}{|w|^{2\D}} + \mO(x)\; \del_\mu \mO(x) \int_a^{a'} d^dw \frac{w^\mu}{|w|^{2\D}} \right.\nonumber\\
 \left.+ \frac{1}{2!}  \mO(x) \; \del_\mu \del_\nu \mO(x) \int_a^{a'} d^dw \frac{w^\mu w^\nu}{|w|^{2\D}} + ...\right)
\end{align}
Using the result \eqref{eq:app-integral} in \autoref{app:results},
\begin{align}
    =& \frac{\fd_0^2}{2} \left( \dfrac{2\pi^{d/2}}{\Gamma\left(\frac{d}{2}\right)} \right) \left( \dfrac{a'^{d-2\D}-a^{d-2\D}}{d-2\D} \right) \left( \int d^dx\; \mO^2(x) \right)  \nonumber\\
    &+ \frac{\fd_0^2}{2}  \left( \dfrac{\pi^{d/2}}{2\ \Gamma\left(\frac{d}{2}+1\right)} \right) \left( \dfrac{a'^{d-2\D+2}-a^{d-2\D+2}}{d-2\D+2} \right) \left( \int d^dx\; \mO \, (\del^2) \mO(x) \right) + \ldots
\end{align}
We see that derivative double-trace couplings are
automatically generated.
The couplings at the new cut-off $a'$ are then,
\begin{equation}\begin{aligned}\label{new-couplings-f0-only}
      \fd'_0 &= \fd_0 - \fd_0^2  \left( \dfrac{2\pi^{d/2}}{\Gamma\left(\frac{d}{2}\right)} \right) \left( \dfrac{a'^{d-2\D}-a^{d-2\D}}{d-2\D} \right) + \ldots\\
      \fd'_1 &= - \fd_0^2  \left( \dfrac{\pi^{d/2}}{2\ \Gamma\left(\frac{d}{2}+1\right)} \right) \left( \dfrac{a'^{d-2\D+2}-a^{d-2\D+2}}{d-2\D+2} \right) + \ldots\\
      \fd'_2 &= - \fd_0^2   \left( \dfrac{\pi^{d/2}}{16\ \Gamma\left(\frac{d}{2}+2\right)} \right) \left( \dfrac{a'^{d-2\D+4}-a^{d-2\D+4}}{d-2\D+4} \right)  + \ldots \footnotemark
\end{aligned}\end{equation}
\footnotetext{Recall, we had started with only $\fd_0 \neq 0$, rest all $\fd_i=0 \forall i>0$.}
The ellipsis in the above equations denotes higher order terms coming from ellipsis in \eq{ft-partition}. \eq{new-couplings-f0-only} can be used to compute $\b$-functions. The contributions coming from terms in ellipsis above are $\sim (\d a)^2$ and hence don't contribute to $\b$-function computations.
\begin{subequations}\begin{align}\label{eq:beta-dimensional-f0-only}
                     \b^{(d)}_0 &= \lim_{a\to a'} \left( a \cdot \dfrac{\fd'_0-\fd_0}{a'-a} \right) =  - \fd_0^2 \ a^{d-2\D} \left( \dfrac{2\pi^{d/2}}{\Gamma\left(\frac{d}{2}\right)} \right) \\
                      \b^{(d)}_1 &= \lim_{a\to a'} \left( a \cdot \dfrac{\fd'_1}{a'-a} \right) =  - \fd_0^2 \ a^{d-2\D+2}  \left( \dfrac{\pi^{d/2}}{2\ \Gamma\left(\frac{d}{2}+1\right)} \right) \\
                       \b^{(d)}_2 &= \lim_{a\to a'} \left( a \cdot \dfrac{\fd'_2}{a'-a} \right) =  - \fd_0^2 \ a^{d-2\D+4} \left( \dfrac{\pi^{d/2}}{16\ \Gamma\left(\frac{d}{2}+2\right)} \right) \\
                       \vdots \nonumber
\end{align}\end{subequations}
where, $\b^{(d)}$ are the $\b$-functions for the dimensionful couplings. In terms of the dimensionless couplings, for the operators with dimension given by \eqref{relevant}, these become,
\begin{subequations}\begin{align}\label{eq:beta-dimensionless-f0-only}
                     \b_0 &= 2\nu \fnd_0  - \fnd_0^2 \ \left( \dfrac{2\pi^{d/2}}{\Gamma\left(\frac{d}{2}\right)} \right) \\
                      \b_1 &= (2\nu-2) \fnd_1 - \fnd_0^2 \ \left( \dfrac{\pi^{d/2}}{2\ \Gamma\left(\frac{d}{2}+1\right)} \right) \\
                       \b_2 &= (2\nu-4) \fnd_2  - \fnd_0^2 \ \left( \dfrac{\pi^{d/2}}{16\ \Gamma\left(\frac{d}{2}+2\right)} \right) \\
                       \vdots \nonumber
\end{align}\end{subequations}
More generally, we can start with double-trace couplings with
arbitrary number of derivatives as in \eq{all-double-trace}. By a simple
generalisation of the above method, we get a closed set of
beta-functions. This is what we describe in what follows.

\subsection[\texorpdfstring{$\b$-function of a general coupling with arbitrary cut-off regulator}{Beta-function of a general coupling with arbitrary cut-off regulator}]{$\b$-function of a general coupling with arbitrary cut-off 
regulator\label{details-FT}} In this section we generalise the
above computations of the $\b$-functions to couplings constants of the double-trace operators with derivatives. The fixed point Lagrangian is perturbed
by a term as follows,
\begin{equation}
    \half \int d^dx \left(\fd_0 \mO^2(x) + \fd_1 \mO\del^2\mO(x) +
    \fd_2 \mO \del^4\mO(x) + \cdots\right)
\end{equation}
where $\fd_i$ are the dimensionful coupling constants for the
operators of the type $\mO (\del^2)^i \mO(x)$, same as in \eq{all-double-trace}, but written in a concise notation. These are the same
class of operators for which $\b$-functions were computed in bulk in
\cite{Heemskerk:2010hk, Faulkner:2010jy}. In a large $N$ theory, the anomalous dimension of the
double-trace operators are suppressed by $1/N$, and so the conformal
dimension of any of the above double-trace operators is $\D_i =
\left[\mO(\del^2)^i \mO(x)\right] = d-2\nu+2i$.  \footnote{We only require $\D_i = \D_{\mO^2}+2i$ in most of our analysis, using the specific value only in $\b$-function computations.}
We are considering appropriately orthogonalized single-trace operators at the
fixed point such that under RG only the multi-traces and their derivatives are generated. We package the above couplings into a single function of $\del^2$ (or equivalently $k^2$ in momentum space),
\begin{equation}\label{eq:packaged-coupling-FT}
    \fd(\del^2) = \fd_0 + \fd_1 (\del^2) + \fd_2 (\del^2)^2 + \cdots
\end{equation}
and hence the double-trace perturbations become,
\begin{equation}\label{eq:perturbation-Langrangian-FT}
    \mL_{DT} = \half \int d^dx\  \mO \ \fd(\del^2) \mO(x)
\end{equation}
In a large-$N$ theory, all the $O(1)$ connected diagrams factorise through the double-trace vertices into chain-like diagrams,
\begin{figure}[H]
    \centering
    \includegraphics[scale=0.5,keepaspectratio=true]{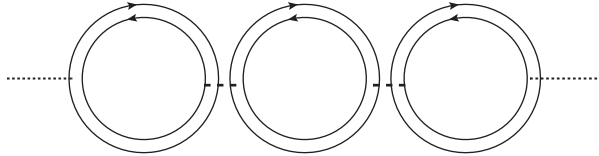}
    \caption{Factorisation through double-trace vertices in Large-$N$ limit.}
    \label{fig:chain-factorization}
\end{figure}
In the \autoref{fig:chain-factorization}, each circle is representative of $\lan\mO(x_1) \mO(x_2)\ran$ contractions, or of their derivatives coming from the double-trace vertices (although here it looks like $\mO = Tr[\Phi^2]$, it is representative of any arbitrary single-trace operator). In a regulated theory a UV cut-off modifies the short-distance behaviour of any correlator. We capture the effect of such regulations in our correlators by introducing a regulating-function, $\mK(|x_1-x_2|/a)$, such that the new regulated correlator becomes, 
\begin{align}
G_a(x_1-x_2) =\lan \mO(x_1) \mO(x_2) \ran_a = 
\dfrac{\mK(|x_1-x_2|/a)}{|x_1-x_2|^{2\D}}. 
\label{regulator}
\end{align}
Here $a$ parametrises the length-scale of regulation, and the correlator shows deviation from polynomial law only near length-scales $\lesssim a$, while long distance behaviour remains power-law, as governed by conformal symmetry. Thus, $\mK(|x_1-x_2|/a) \rightarrow 1$, when $|x_1-x_2|\gg a$, but falls off faster than $|x_1-x_2|^{2\D}$, when $|x_1-x_2| \lesssim a$. In our study, we assume that the short-distance fall-off is fast enough to regulate all the correlators $\lan (\del^2)^i\mO(x_1) \ (\del^2)^j \mO(x_2) \ran$ at short distances. An example of such a regulator is $\mK(r/a) = \Theta(r-a)$, where $\Theta$ is the Heaviside-theta function, which was used in \cite{Vecchi:2010jz,Witten:2001ua,Vecchi:2010dd}. We also use a regulated form of $\Theta$-function,
\begin{align}\label{eq:Regulated-Theta}
	\mK(\rho) = \frac{\sqrt{\pi } e^{1/\omega ^2} \left(\omega ^2+2\right) \Big( \text{erf}\left(\frac{\rho -1}{\omega }\right)+\text{erf}\left(\frac{1}{\omega }\right) \Big) + 2 \omega - 2 (\rho +1) \omega  e^{-\left( \rho ^2-2 \rho \right) / \omega ^2}}{\sqrt{\pi } e^{1/\omega ^2} \left(\omega ^2+2\right) \Big(\text{erf}\left(\frac{1}{\omega }\right)+1 \Big)+2 \omega }
\end{align}
The corresponding regulated $\d$-function that is 
$$ \d_r(\rho-1) = \frac{4 \rho ^2 e^{-\frac{(\rho -2) \rho }{\omega ^2}}}{\omega  \left(\sqrt{\pi } e^{\frac{1}{\omega ^2}} \left(\omega ^2+2\right) \left(\text{erf}\left(\frac{1}{\omega }\right)+1\right)+2 \omega \right)}
$$
here, $\omega$ is the width of the regulated $\d$-function and regulated $\Theta$-function.\\
Hence, computation of any physical observable involves evaluation of chain-diagrams with regulated correlators.

Evaluation of $\b$-functions involves studying the change of the coupling constants $\fd_i$ under the change of the cut-off scale $a \rightarrow a'$. All the physical observables in this new theory are required to remain unchanged and the chain diagrams involve the correlators, $G_{a'}(|x_1-x_2|/a')$. One can relate the diagrams in the original theory at $a$ to those in the new theory at cut-off $a'$ by relating the correlators.
\begin{equation}\label{eq:PropA-PropAp}
    G_a(x_1-x_2) = G_{a'}(x_1-x_2) + \del_{a'} G_{a'}(x_1-x_2) \ (-\d a) + \ldots
\end{equation}
Note that the second term above involves derivative of $\mK(|x-y|/a)$ and is supported only in the region $|x-y|\sim a'$. The first term on the RHS of \eqref{eq:PropA-PropAp} contributes to the chain-diagrams at the new cut-off $a'$ and subsequent terms correct the coupling constant. Integration involving second and subsequent terms can be seen as coming from integration of \emph{heavy modes}, as they contribute only at short distances. We will denote them by coloured contractions in our diagrammatic representations, as in \autoref{fig:pert_heavy}.

\begin{figure}[t!]
    \centering
    \includegraphics[scale=0.4,keepaspectratio=true]{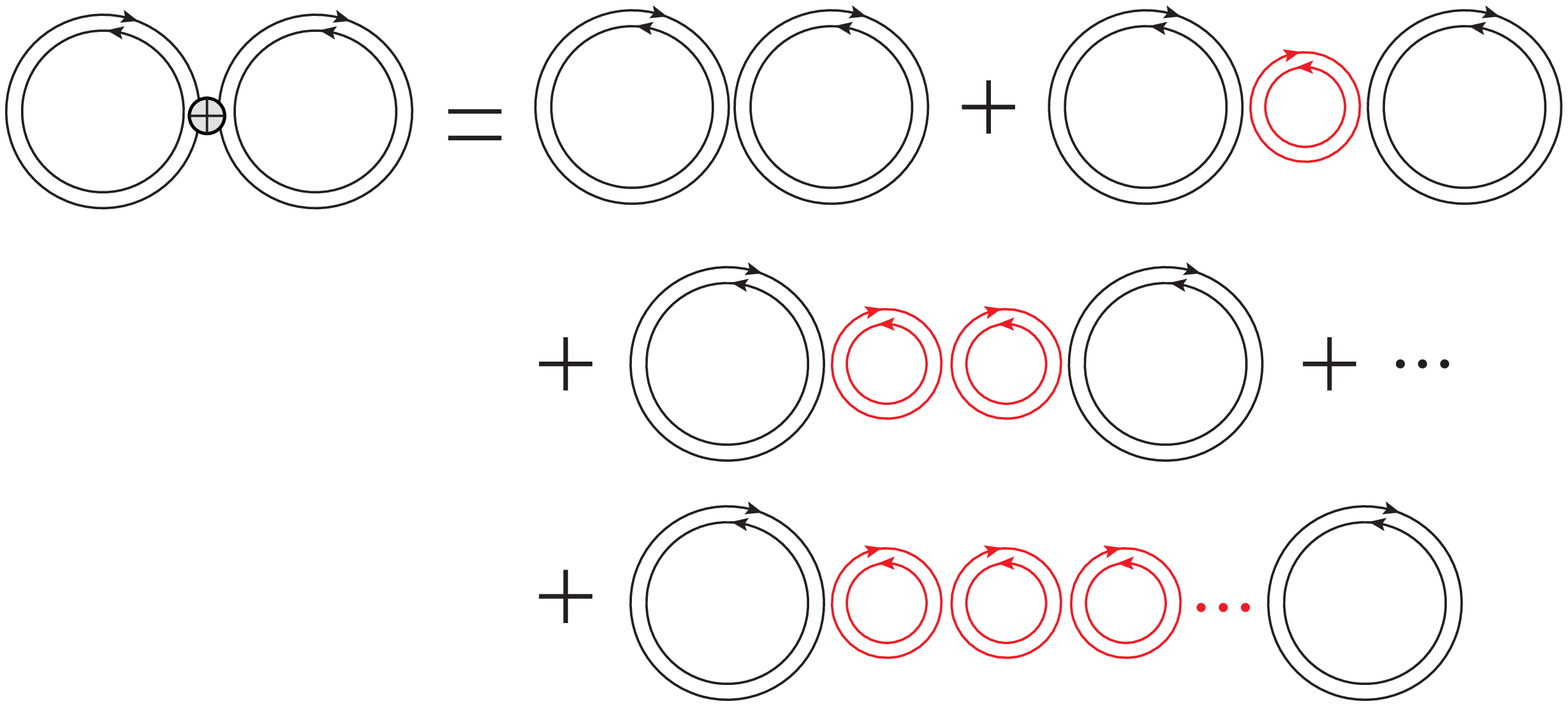}
    \caption{Corrections to a vertex at new cut-off $a'$. The crossed vertex on LHS denotes the vertex at new cut-off. Vertices on RHS are original vertices at $a$. Coloured contractions denote integration of heavy modes coming from higher order corrections in \eqref{eq:PropA-PropAp}.}
    \label{fig:pert_heavy}
\end{figure}

We compute the contribution of the second diagram on the RHS of \autoref{fig:pert_heavy} with the vertices $\half \fd_n \int d^dz_1 \mO \ (\del^2)^n \mO(z_1)$  and $\half \fd_m \int d^dz_2 \mO \ (\del^2)^m \mO(z_2)$. There are 4 ways to choose the \emph{heavy contractions} between single-trace operators,
\begin{align}\label{eq:vertex-correction}
    & \frac{\fd_n \fd_m}{4} (-\d a') \left( \int d^dz_1 d^dz_2 \ \mO(z_1) \ \del_{a'}\left[(\del^2)^n G_{a'} (z_1-z_2)\right] (\del^2)^m\mO(z_2) \right. \nonumber \\
    &+ \int d^dz_1 d^dz_2 \ \mO(z_1) \ \del_{a'}\left[(\del^2)^{m+n} G_{a'} (z_1-z_2)\right] \mO(z_2) \nonumber \\
    &+ \int d^dz_1 d^dz_2 \ (\del^2)^n \mO(z_1) \ \del_{a'}\left[ G_{a'} (z_1-z_2)\right] (\del^2)^m\mO(z_2) \nonumber \\
    & \left. + \int d^dz_1 d^dz_2 \ (\del^2)^n \mO(z_1) \ \del_{a'}\left[(\del^2)^m G_{a'} (z_1-z_2)\right] \mO(z_2) \right) 
\end{align}
Here we have kept only the linear variation in $(\d a')$, since only that is required in the $\b$-function computations. All the subsequent terms in \eqref{eq:PropA-PropAp} (which are higher order in $(\d a')$) don't contribute to the $\b$-functions, even though they need to be considered in computation of the exact vertex at the new cut-off. For the same reason second and following rows in \autoref{fig:pert_heavy} don't contribute to the $\b$-function computation. As in any differential equation, their contribution is exactly captured in the solution. At this point the $\b$-functions in large-$N$ limit are quadratic, whose exactness will be established in \autoref{subsec:exactness}. This is consistent with the holographic computations of the $\b$-functions.\\
In \eqref{eq:vertex-correction}, we can write the operator at $\mO(z_2)$ in a Taylor series expansion around $z_1$.
\begin{align}
    (\del^2)^m\mO(z_2) = (\del^2)^m\mO(z_1) &+ (z_2-z_1)^\mu \del_\mu \Big( (\del^2)^m\mO(z_1) \Big) \nonumber \\
    &+ \frac{1}{2!} (z_2-z_1)^\mu (z_2-z_1)^\nu \del_\mu \del_\nu \Big( (\del^2)^m\mO(z_1) \Big) + \cdots
\end{align}

From the conformal field theory point of few, this is same as translating the operator at $z_1$ to $z_2$. Furthermore, rotational invariance of the theory implies that only the vector-singlets constructed at any level of Taylor series contribute, and hence odd-terms in the Taylor series don't contribute. Thus a general term appearing in the Taylor series can be written as,
\begin{align}\label{eq:quadratic-term-in-beta}
    \int d^dz_1 d^dz_2 \mO(z_1) \del_{a'}\left[ (\del^2)^n G_{a'}(|z_1-z_2|)\right] \Bigg(\frac{1}{(2k)!}  z_{21}^{\mu_1}\ldots z_{21}^{\mu_{2k}} \del_{\mu_1}\ldots\del_{\mu_{2k}} \Big( (\del^2)^m\mO(z_1) \Big) \Bigg) \nonumber \\
    =(a')^{2k} \left( \frac{2^{1-2k} \pi^{d/2}}{\Gamma(k+1) \Gamma(k+\frac{d}{2})} \right) \times \left( \int d\rho \ \rho^{d-1+2k} \ \del_{a'}\left[(a')^{-2n} (\del_\rho^2)^n G_{a'}(a' \rho)\right] \right) \nonumber \\
    \times \ \int d^dz_1 \mO(z_1) (\del^2)^{(m+k)}\mO(z_1)
\end{align}
where, we have used the notation $\vec{\rho} = \dfrac{\vec{z}_{21}}{a'}$, $z_{ij}= z_i-z_j$, $\rho=|\vec{\rho}|$; and the first numerical factor is coming from the angular integrations (see \autoref{app:results}).\\
Clearly, $\b$-function of every coupling constant in the double-trace perturbation, $\fd_i$, is quadratic in every other coupling constant, $\fd_j$. It is instructive to note that the contribution of some coupling $f_n$ to the $\b_i$, where $n>i$ comes only from those terms in \eqref{eq:vertex-correction} in which the operator $(\del^2)^n\mO$ is involved in a contraction.\\
We show here only first few $\b$-functions, while we have pushed the details of calculations to \autoref{app:results}:
\begin{subequations}\label{eq:General-Beta}\begin{align}
    \b_0 &= 2 \nu \fnd_0 + \fnd_0^2 \left( \a_0 \Gcoff^{\mK'}_\D \right) + \fnd_0 \fnd_1 \ \a_0 \left[ \rho^{d-2\D-1} \left( \rho \ \mK^{(2)}(\rho) - (2\D-1) \mK^{(1)}(\rho) \right) \right]_0^\infty  \nonumber \\
    &+ \fnd_1^2 \ \a_0 \left[ \rho^{d-2\D-3} \left(\rho^3 \mK^{(4)}(\rho) - \rho^2 (6\D-d-2) \mK^{(3)}(\rho) \right. \right. \nonumber\\
    &\hspace{2cm}+ \rho \left( 12 \Delta ^2-(4 d+2) \Delta +d -1 \right) \mK^{(2)}(\rho) \nonumber \\
    &\hspace{2cm}- \left. \left. \left(4 \Delta ^2-1\right) (2 \Delta -d+1)\mK^{(1)}(\rho) \right) \right]_0^\infty \\[15pt]
    \b_1 &= (2 \nu-2) \fnd_1 + \fnd_0^2 \left( \a_1 \Gcoff^{\mK'}_{\D-1} \right) \nonumber \\
    &+ \fnd_0 \fnd_1 \left( (\a_0 + 2 d \ \a_1) \ \Gcoff^{\mK'}_{\D} + \a_1  \ \left[ \rho^{d-2\D+1} \left( \rho \mK^{(2)}(\rho) - (2\D+1) \mK^{(1)}(\rho) \right) \right]_0^\infty \right)  \nonumber \\
    &+ \fnd_1^2 \Bigg( \a_1 \ \left[ \rho^{d-2\D-1} \left( \rho^3 \mK^{(4)}(\rho) - \rho^2 (6\D-d) \mK^{(3)}(\rho) \right. \right. \nonumber\\
    &\hspace{2cm}+ \rho \left( 12 \Delta ^2-(4 d+6) \Delta +d -3 \right) \mK^{(2)}(\rho) \nonumber \\
    &\hspace{2cm}- \left. \left. \left(4 \Delta ^2-1\right) (2 \Delta -d+3)\mK^{(1)}(\rho) \right) \right]_0^\infty \nonumber \\
    &\hspace{1cm}+ 2 \ \a_0 \ \left[ \rho^{d-2\D-1} \left( \rho \ \mK^{(2)}(\rho) - (2\D-1) \mK^{(1)}(\rho) \right) \right]_0^\infty \Bigg)
\end{align}\end{subequations}
where, we have used the following short-hand notations 
\begin{equation}\begin{aligned}\label{eq:short-hands}
    \a_i &= \frac{2^{1-2i} \ \pi^{d/2}}{\G(i+1) \G(i+\half d)} \\[5pt]
    \Gcoff^{\mK^{(n)}}_{\D} &= \int d\rho \ \rho^{d-2\D} \ \mK^{(n)}(\rho)
\end{aligned}\end{equation}
$\fnd_i = a^{2\nu-2i} \fd_i$ are the dimensionless coupling constants for the operator $\mO \: (\del^2)^i \mO(x)$ (corresponding to choice \eqref{relevant}), and the $\b$-functions are computed for these dimensionless couplings.

It is apparent that some of the coefficients in the above $\b$-functions are simply boundary terms. With our assumption that the regulation scheme, $\mK$ falls off fast enough at the origin to regulate all the correlators, these coefficients vanish. Thus the $\b$-functions become,
\begin{subequations}\label{eq:General-Beta-W/O-BDY}\begin{align}
    \b_0 &= 2 \nu \fnd_0 - \fnd_0^2 \left( \a_0 \Gcoff^{\mK'}_\D \right) \\[5pt]
    \b_1 &= (2 \nu-2) \fnd_1 - \fnd_0^2 \left( \a_1 \Gcoff^{\mK'}_{\D-1} \right) - \fnd_0 \fnd_1 \left( (\a_0 + 2 d \ \a_1) \ \Gcoff^{\mK'}_{\D} \right)
\end{align}\end{subequations}
We find that the $\b$-functions follow a pattern in which the coefficient of $\fnd_i \fnd_j$ in $\b_k$ is only a boundary term when $i+j>k$, and hence vanish. We have checked it explicitly for first four $\b$-functions listed below and could easily see it generalise to any arbitrary order,
\begin{equation}\begin{aligned}\label{eq:Beta-FT-final}
    \b_0 &= 2 \nu \fnd_0 - \fnd_0^2 \left( \a_0 \Gcoff^{\mK'}_\D \right) \\[10pt]
    \b_1 &= (2 \nu-2) \fnd_1 - \fnd_0^2 \left( \a_1 \Gcoff^{\mK'}_{\D-1} \right) - \fnd_0 \fnd_1 \left( (\a_0 + 2 d \ \a_1) \ \Gcoff^{\mK'}_{\D} \right) \\[5pt]
    &= (2 \nu-2) \fnd_1 - \fnd_0^2 \left( \a_1 \Gcoff^{\mK'}_{\D-1} \right) - 2\fnd_0 \fnd_1 \left( \a_0 \ \Gcoff^{\mK'}_{\D} \right) \\[10pt]
    \b_2 &= (2 \nu-4) \fnd_2 - \fnd_0^2 \left( \a_2 \Gcoff^{\mK'}_{\D-2} \right) - \fnd_0 \fnd_1 \left( (\a_1 + 4(d+2)\a_2) \Gcoff^{\mK'}_{\D-1} \right) \\
    &\hspace{1.5cm} - \fnd_0 \fnd_2 \left( (\a_0 + 8 d (d+2) \a_2) \Gcoff^{\mK'}_\D \right) - \fnd_1^2 \ \frac{1}{4} \left(   (\a_0 + 4d \a_1 + 8d (d+2) \a_2) \Gcoff^{\mK'}_{\D} \right) \\[5pt]
    &= (2 \nu-4) \fnd_2 - \fnd_0^2 \left( \a_2 \Gcoff^{\mK'}_{\D-2} \right) - 2 \fnd_0 \fnd_1 \left( \a_1 \Gcoff^{\mK'}_{\D-1} \right) - \left( 2 \fnd_0 \fnd_2 + \fnd_1^2 \right) \left( \a_0 \Gcoff^{\mK'}_\D \right) \\[10pt]
    \b_3 &= (2 \nu-6) \fnd_1 - \fnd_0^2 \left( \a_3 \Gcoff^{\mK'}_{\D-3} \right) - 2 \fnd_0 \fnd_1 \left( \a_2 \Gcoff^{\mK'}_{\D-2} \right) - \left( 2 \fnd_0 \fnd_2 + \fnd_1^2 \right) \left( \a_1 \Gcoff^{\mK'}_{\D-1} \right) \\
    &\hspace{8.3cm} - 2 \left( \fnd_0 \fnd_3 + \fnd_1 \fnd_2 \right) \left( \a_0 \Gcoff^{\mK'}_\D \right) \\
    \vdots
\end{aligned}\end{equation}
We have used the identity $\a_i = (2i+2)(d+2i)\a_{i+1}$ to simplify coefficients, and $\a_i$ and $\Gcoff^{\mK'}_{\Delta}$ are given by \eqref{eq:short-hands}. In \autoref{tab:Couplings} we summarise the values of the coefficients above for $\mK = \Theta$ and $\mK = $\eqref{eq:Regulated-Theta}, the regulated $\Theta$-function.
\begin{table}
\begin{center}
\begin{tabular}[h]{c|c|c}
 \toprule[1.2pt]
 & $\Theta(\rho)$ & Regulated-$\Theta(\rho)$ \\[5pt]
 \midrule
 $\a_0 \Gcoff^{\mK'}_{\D}$ & $\dfrac{2\pi^{d/2}}{\Gamma(\frac{d}{2})}$ & $ \frac{\substack{ 4 \pi ^{d/2} \omega ^{2 \nu +1} \left[ 2 \Gamma (\nu +2) \, _1F_1\left(\nu +2;\frac{3}{2};\frac{1}{\omega ^2}\right) \right. \\ \hspace*{2cm} \left.+\omega  \Gamma \left(\nu +\frac{3}{2}\right) \, _1F_1\left(\nu +\frac{3}{2};\frac{1}{2};\frac{1}{\omega ^2}\right)\right]}}{\Gamma \left(\frac{d}{2}\right) \Big[ \sqrt{\pi } e^{\frac{1}{\omega ^2}} \left(\omega ^2+2\right) \times \left(\text{erf}\left(\frac{1}{\omega }\right)+1\right) +2 \omega \Big]} $ \\[20pt]
 $\a_1 \Gcoff^{\mK'}_{\D-1}$ & $\dfrac{\pi^{d/2}}{2\Gamma(\frac{d}{2}+1)}$ & $ \frac{\substack{\pi ^{d/2} \omega ^{2 \nu +3} \left[ 2 \Gamma (\nu +3) \, _1F_1\left(\nu +3;\frac{3}{2};\frac{1}{\omega ^2}\right) \right. \\ \hspace*{2cm} \left.+\omega  \Gamma \left(\nu +\frac{5}{2}\right) \, _1F_1\left(\nu +\frac{5}{2};\frac{1}{2};\frac{1}{\omega ^2}\right)\right]}}{\Gamma \left(\frac{d}{2}+1\right) \Big[ \sqrt{\pi } e^{\frac{1}{\omega ^2}} \left(\omega ^2+2\right) \times \left(\text{erf}\left(\frac{1}{\omega }\right)+1\right) +2 \omega \Big]} $ \\[20pt]
 $\a_2 \Gcoff^{\mK'}_{\D-2}$ & $\dfrac{\pi^{d/2}}{16\Gamma(\frac{d}{2}+2)}$ & $ \frac{\substack{\pi ^{d/2} \omega ^{2 \nu +5} \left[ 2 \Gamma (\nu +4) \, _1F_1\left(\nu +4;\frac{3}{2};\frac{1}{\omega ^2}\right) \right. \\ \hspace*{2cm} \left.+\omega  \Gamma \left(\nu +\frac{7}{2}\right) \, _1F_1\left(\nu +\frac{7}{2};\frac{1}{2};\frac{1}{\omega ^2}\right)\right]}}{8 \Gamma \left(\frac{d}{2}+2\right) \Big[ \sqrt{\pi } e^{\frac{1}{\omega ^2}} \left(\omega ^2+2\right) \times \left(\text{erf}\left(\frac{1}{\omega }\right)+1\right) +2 \omega \Big]} $\\[20pt]
 $\a_3 \Gcoff^{\mK'}_{\D-3}$ & $\dfrac{\pi^{d/2}}{192 \Gamma(\frac{d}{2}+3)}$ & $ \frac{\substack{\pi ^{d/2} \omega ^{2 \nu +7} \left[ 2 \Gamma (\nu +5) \, _1F_1\left(\nu +5;\frac{3}{2};\frac{1}{\omega ^2}\right) \right. \\ \hspace*{2cm} \left.+\omega  \Gamma \left(\nu +\frac{9}{2}\right) \, _1F_1\left(\nu +\frac{9}{2};\frac{1}{2};\frac{1}{\omega ^2}\right)\right]}}{96 \Gamma \left(\frac{d}{2}+3\right) \Big[ \sqrt{\pi } e^{\frac{1}{\omega ^2}} \left(\omega ^2+2\right) \times \left(\text{erf}\left(\frac{1}{\omega }\right)+1\right) +2 \omega \Big]} $ \\[20pt]
 \bottomrule[1.5pt]
\end{tabular}
\caption{List of the coefficients appearing in \eqref{eq:Beta-FT-final} for choice of two different regulators discussed in the text.}
\label{tab:Couplings}
\end{center}
\end{table}

\subsection[\texorpdfstring{Exactness of $\b$-function}{Exactness of beta-function}]{Exactness of $\b$-function}\label{subsec:exactness}
The usual Wilsonian or Polchinski-Wilsonian renormalization procedure involves integration of UV/short-distance-degrees of freedom. In a continuum field theory defined around Gaussian fixed point, momentum eigenvalues serve as adequate label to differentiate between UV and IR degrees of freedom, and \emph{heavy modes} are defined as those modes with momentum greater than some arbitrary cut-off value. When we change the value of the cut-off, those modes that lie between the old and new cut-offs are integrated over. Diagrammatically these are denoted by bold lines, and in this paper they are represented by coloured lines (see \autoref{fig:wilson}). In this paper, we perform an integration of \emph{heavy modes} in position space, as demonstrated above and we justify our approach in this subsection.
\begin{figure}[H]
    \centering
    \includegraphics[scale=0.75,keepaspectratio=true]{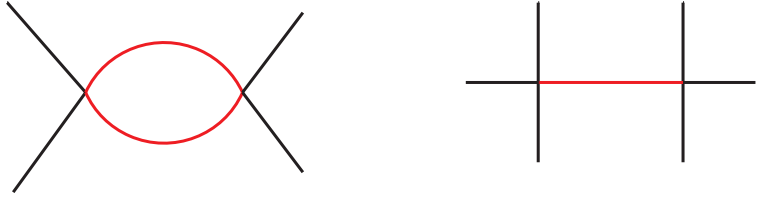}
    \caption{Types of diagrams that originate in Wilsonian RG due to integration of heavy modes. Coloured lines represent heavy modes that are being integrated out. Above diagrams show the origin of corrections to $\phi^4$ and $\phi^6$ vertices.}
    \label{fig:wilson}
\end{figure}

\begin{figure}[!ht]
    \centering
    \includegraphics[scale=0.4,keepaspectratio=true]{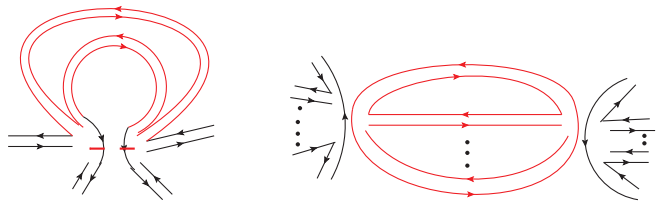}
    \caption{Diagrams that arise in contraction of \emph{heavy modes} in a matrix theory from the double-trace vertices. The first kind of diagrams correct the single-trace coupling constants at sub-leading order of $N$ counting. Only the second kind of diagrams correct the double-trace coupling constants, at the leading order.}
    \label{fig:double-trace-contractions}
\end{figure}

In a large-$N$ matrix theory like the one that we are considering, integration of \emph{heavy modes} generates diagrams shown in \autoref{fig:double-trace-contractions}.
With our normalisation of operators, it is clear that the leading contribution comes from contracting all the \emph{heavy `legs'} between two double-trace vertices, so one effectively has $\lan \mO(x_1) \mO(x_2) \ran$. Fewer contractions of legs leaves us with one more propagator (with a contribution of $1/N$) than number of loops (which contribute a factor of $N$ each), and hence the contribution is suppressed. Moreover, such a diagram with fewer \emph{heavy contractions} contribute to a triple trace term, which even though comes with the correct normalisation (of $1/N$) in our N counting, doesn't contribute to $O(1)$ part of the effective action.

\begin{figure}[!ht]
    \centering
    \includegraphics[scale=0.5,keepaspectratio=true]{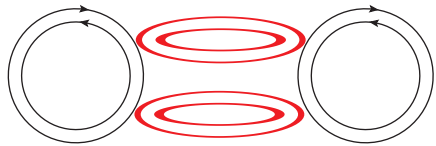}
    \caption{An example of a 2-loop diagram that is suppressed in large-$N$ counting. Suppression of similar diagrams is also discussed in \cite{Pomoni:2008de}.}
    \label{fig:2-loop-suppressed}
\end{figure}

There is a class of diagrams as shown in \autoref{fig:2-loop-suppressed}, which are suppressed by appearance of internal propagators. In general, any diagram that involves internal propagators are suppressed. A similar reasoning appears in \cite{Pomoni:2008de} in terms of certain auxiliary fields that are used to write the double-trace operators in terms of the single-trace operators. Thus, it is clear that the only diagrams that can possibly contribute at the leading order are the chain-type diagrams discussed previously in this section, and hence the $\b$-functions computed using such diagrams are exact.

\subsection{Field theory correlators in momentum space}\label{sub-sec:ft-corr-mom-sp}
Most of our computations in bulk are in momentum space. For sake of completeness and to be able to compare the results, we will summarize some of the field theory results in momentum space. The momentum space expression for the field theory correlator along with the inclusion of the regulating function, \eqref{regulator}, in general is of the form,
\begin{equation}\label{gen-regulatedcorr}
	\lan \mO(k)\mO(-k) \ran_\e = p^{2\D-d} + \e^{d-2\D} \left(a_0 + a_1(p\e)^2 + a_2(p\e)^4 + \ldots \right)
\end{equation}
where, $a_i$ are some coefficients that are given by the choice of the regulating function $\mK$. For example, for the $\t$-function regulation, we have following correlator in momentum space (to keep in line with the bulk notations, we are using $\D = d/2 \pm \nu$),
\begin{align}\label{th-regulated-corr-mom}
	\lan \mO(k)\mO(-k) \ran_\e & = k^{\pm 2 \nu } \left(-4 \pi ^{\frac{d-1}{2}} \cos (\pi  \nu ) \dfrac{\Gamma (\mp 2 \nu -1)}{\Gamma \left(\dfrac{d-1}{2}\right)}\right) \pm 2 \e^{\mp2 \nu } \, \pi ^{\frac{d-1}{2}} \frac{ _1F_2\left(\mp\nu ;\frac{3}{2},\mp\nu +1;-\frac{1}{4} (k \e)^2 \right)}{\nu\ \Gamma \left(\frac{d-1}{2}\right) }\nonumber\\
	&= k^{\pm 2 \nu } \left(-4 \pi ^{\frac{d-1}{2}} \cos (\pi  \nu ) \dfrac{\Gamma (\mp 2 \nu -1)}{\Gamma \left(\dfrac{d-1}{2}\right)}\right) \nonumber \\
      &\hspace{13pt} + \e^{\mp 2 \nu }  \left( \pm \frac{2 \pi ^{\frac{d-1}{2}} }{\nu \  \Gamma \left(\dfrac{d-1}{2}\right)} \right) \left[ 1 \pm \nu \frac{(k \e)^2}{6 ( \nu + 1)} \mp \nu \frac{(k \e)^4}{120 (\nu +2)} \pm \nu \frac{(k \e)^6}{5040 (\nu +3)} + \ldots \right]
\end{align}
and the coefficients $a_i$ can be read from the above equation. Strictly speaking, in the correctly regulated IR theory, we don't get the diverging counter terms in the above correlators. That is to say, for example, if $0<\nu<1$, then around the IR fixed point, when $\D=d/2+\nu$, in a correctly regulated theory, the first counter term above, $a_0=0$. ( i.e. we need to add a counter-term with $-a_0$).

In a more general case, it might happen that the kinematic term (the term proportional to $k^{2\D-d}$ in the above equation) also has a multiplicative integer power series in $k\e$. We attribute such a series to a multiplicative wavefunction renormalization of the operator $\mO$. Thus, for any choice of a regulator the 2-point function in momentum space can be brought to the above form. For reference, we have presented the correlator computations in a large N bosonic vector model in \autoref{app:wilson-fisher}. There the correlator for the $\phi^2$ operator in the regulated UV theory is given by, \eqref{bubble} which has the same form as presented above.

In a double-trace deformed field theory around a fixed point, the correlator of the $\mO$ operator in the large N limit is given by the Schwinger-Dyson series,
\begin{align}\label{dyson-schwinger-ft}
	\lan \mO(k)\mO(-k) \ran^{\fd}_\e &= \lan \mO(k)\mO(-k) \ran_\e - \fd(k^2)  \lan \mO(k)\mO(-k) \ran^2_\e + \fd^2(k^2) \lan \mO(k)\mO(-k) \ran^3_\e + \cdots\nonumber\\
	&= \dfrac{\lan \mO(k)\mO(-k) \ran_\e}{1+ \fd(k^2)\lan \mO(k)\mO(-k) \ran_\e}
\end{align}

\paragraph{IR fixed point from UV theory} Now we will analyse the UV and IR limit of the perturbed correlators around the fixed points of the theory. Around the UV fixed point $\D=d/2-\nu$, and the dimensionless coupling constants are $\fnd_-(k\e)=\e^{2\nu}\fd_-(k^2)$, so the perturbed correlator is given by,
\begin{align}\label{pert-corr-uv}
	\lan \mO(k)\mO(-k) \ran^{\fd_-}_\e = \frac{k^{-2\nu} + \e^{2\nu}\d C(k\e)}{ 1 + \e^{-2\nu} \fnd_-  \left( k^{-2\nu} + \e^{2\nu}\d C(k\e)\right)}
\end{align}
Taking the IR limit of this correlator, $k\e\to0$, we get the following limit of the correlator,
\begin{equation}
	\lim_{k\e\to0}\lan\mO(k)\mO(-k)\ran^{\fd_-}_{\e} \to \left( \frac{\e^{2\nu}}{\fnd_-} - k^{2\nu} \frac{\e^{4\nu}}{\fnd_-^2} + k^{4\nu} \frac{\e^{6\nu}}{\fnd_-^3} \left( 1 + \fnd_- \d C\right) + \cdots \right)
\end{equation}
Thus in the strict IR limit, only the second term survives, and in that case we get the correlator of the IR theory upto some wavefunction renormalization, $\e^{4\nu}{\fnd^2}$, and the first contact term, after the inclusion of this wavefunction renormalization becomes, $\fnd \cdot \e^{-2\nu}$,
\begin{equation}
	\lim_{k\e\to0}\lan \tilde{\mO}(k) \tilde{\mO}(-k)\ran^{\fd^*_-}_{\e} \to  \Big( \fnd^*_- \cdot \e^{-2\nu} - p^{2\nu}  \Big)
\end{equation}
In this limit, even the coupling constants approach their respective IR fixed point value, $\fnd \to \fnd^*_-$. So the first term is precisely the type of contact term that one expects for the regulated theory with the scaling dimension, $\D=d/2+\nu$.

\paragraph{UV fixed point from IR theory} Let us analyse the correlator for a double-trace deformed theory around the IR fixed point, and take the UV limit of such a correlator. The correlator given by the exact summation of the Schwinger-Dyson sum in this case is also \eqref{dyson-schwinger-ft}, but now with the correlators at the IR fixed point, and also the perturbation, $\fnd_+({k\e}) = \e^{-2\nu} \fd_+(k^2)$, around this fixed point,
\begin{align}\label{pert-corr-ir}
	\lan \mO(k)\mO(-k) \ran^{\fd_+}_\e = \frac{k^{2\nu} + \e^{-2\nu}\d C(k\e)}{ 1 + \e^{2\nu} \fnd_+ \left( k^{2\nu} + \e^{-2\nu}\d C(k\e)\right)}
\end{align}
The UV limit in this case is, $k\e\to\infty$,
\begin{align}
	\lim_{k\e\to0}\lan\mO(k)\mO(-k)\ran^{\fd_+}_{\e} &\to \left( \frac{\e^{-2\nu}}{\fnd_+} - k^{-2\nu} \frac{\e^{-4\nu}}{\fnd_+^2} + k^{-4\nu} \frac{\e^{-6\nu}}{\fnd_+^3} \left( 1 + \fnd_+ \d C\right) + \cdots \right) \nonumber \\[10pt]
	& \underrightarrow{\substack{\text{on wavefunction}\\ \text{renormalization}}} \quad \Big( \fnd^*_+ \cdot \e^{2\nu} - p^{-2\nu}  \Big)
\end{align}
Thus, we see that starting with either of the fixed points, in correct limits, we can flow to the other fixed point. It is clear that the properties of the correlators and the $\b$-functions that are discussed in this section are also true for the holographic computations. We now discuss a few subtleties that are involved in the duality between the field theory and the bulk.

\section{Scheme-dependence and coupling constant redefinition}\label{sec:rational-fraction}
    In this section we will discuss \emph{(a)} the relationship between the choice of regulator $\mK$ in the field theory and radial cut-off in the holographic computations, and \emph{(b)} how are different choices of regulators $\mK$ related to diffeomorphisms in the space of couplings (or equivalently, in the space of field theories).\\
    In the derivation of $\b$-functions for a general regulator $\mK$, \eqref{eq:Beta-FT-final}, it is clear that all the independent coefficients appearing there are of the form
    \begin{equation}
	  \Gcoff^{\mK'}_{\D-j} = \int d\rho \ \rho^{d-2\D+2j} \ \mK'(\rho), \hspace{1cm} j \in \{\mathbb{Z}^+ \cup 0\}
    \end{equation}
    These are almost like moments of derivative of the regulating function, $\mK$ (we say almost, because $d-2\D=2\nu$ is not an integer). Thus knowledge of all these coefficients, along with the behaviour of $\mK$ at $0$ and $\infty$, is, in principle, enough to reconstruct $\mK$. However, the relationship between the coefficients and the regulating function in the bulk calculation is different, which points to a different `scheme' of renormalization between bulk and field theory.
    
    Next, we will discuss the class of diffeomorphisms in the space of couplings, $\fnd_i$, that correspond to different choices of regulating function in the Wilsonian computation.
    The general structure of the $\b$-functions either in bulk \eqref{eq:st-bulk-indiv-beta-2} and \eqref{eq:aq-bulk-indiv-beta-2} or field theory \eqref{eq:Beta-FT-final}  is:
    \begin{equation}\begin{aligned}\label{eq:general-beta}
	\b_0 &= 2\nu \fnd_0 - \scA_0 \fnd_0^2 \\
	\b_1 &= (2\nu-2) \fnd_1 - \scA_1 \fnd_0^2 - 2 \scA_0 \fnd_0 \fnd_1 \\
	\b_2 &= (2\nu-4) \fnd_2 - \scA_2 \fnd_0^2 - 2 \scA_1 \fnd_0 \fnd_1 - \scA_0 \left( 2 \fnd_0 \fnd_2 + \fnd_1^2\right) \\
	\b_3 &= (2\nu-6) \fnd_3 - \scA_3 \fnd_0^2 - 2 \scA_2 \fnd_0 \fnd_1 - \scA_1 \left( 2 \fnd_0 \fnd_2 + \fnd_1^2\right) - \scA_0 \left( 2 \fnd_1 \fnd_2 + 2 \fnd_0 \fnd_3\right) \\
	\vdots
    \end{aligned}\end{equation}
    for some values of $\scA_i$.\\
    Above $\b$-functions, $\b_i$ and couplings, $f_i$ can be packaged into generating functions defined as
    \begin{subequations}\begin{align}
	\b(\k) &= \b_0 + \k^2 \b_1 + \k^4 \b_2 + \k^6 \b_3 + \cdots \label{eq:packaged-beta-couplings-a} \\
	\fnd(\k) &= \fnd_0 + \k^2 \fnd_1 + \k^4 \fnd_2 + \k^6 \fnd_3 + \cdots \label{eq:packaged-beta-couplings-b}
    \end{align}\end{subequations}
    and then \eqref{eq:general-beta} is re-packaged into a single equation,
    \begin{equation}\label{eq:packaged-beta-eqn}
	\b(\k) = 2 \nu \fnd(\k) - \mathcal{A}(\k) \fnd^2(\k) - \k \del_\k \fnd(\k)
    \end{equation}
    where,
    \begin{equation}\label{eq:A-series}
	\scA(\k) = \scA_0 + \k^2 \scA_1 + \k^4 \scA_2 + \k^6 \scA_3 + \cdots
    \end{equation}
    Note that, with the identification $\k= \e k$ in \eqref{eq:packaged-beta-couplings-b}, we have the dimensionless version of $\bfd(k)=\sum_{n=0}^\infty \bfd_n {(k^2)}^n$ in \eqref{all-double-trace}. Then, \eqref{eq:packaged-beta-eqn} becomes,
    \begin{equation}\label{eq:totally-packaged-beta}
	\dot{\fnd}(\k) = \e \del_\e \fnd(\k)|_k = 2 \nu \fnd(\k) - \mathcal{A}(\k) \fnd^2(\k) \nonumber
    \end{equation}
    Such a packaged form of $\b$-functions appears naturally in the bulk computations (see \eqref{eq:st-bulk-beta-dC} and \eqref{eq:aq-bulk-beta-dC}). \\
    The above differential equation can be rewritten as,
    \begin{equation}
	\left( \frac{\e^{2\nu}}{\fnd(\k)} \right)^{\Cdot[2]} = \e^{2\nu} \scA(\k) \nonumber
    \end{equation}%
    From the field theory computations, we see that different choices of regulating functions, $\mK$, correspond to different $\scA_i$.
    Now, if we have another set of $\b$-function differential equations with different coefficients, packaged into $\scAb(\k)$, which denotes a different \emph{scheme} of renormalization, then between two different set of $\b$-functions, the couplings in these two different schemes, $\fnd(\k)$ and $\bfnd(\k)$ can be related by,\footnote{For our interest, the Wilsonian/Polchinski-Wilsonian scheme and Holographic scheme are the ones that we want to relate, and hence we use the same notations for couplings as those we have used previously in this paper, $\fnd$ for dimensionless field theory couplings, and $\bfnd$ for dimensionless bulk couplings}
    \begin{equation}
	\left( \e^{2\nu} \left( \frac{1}{\fnd(\k)} - \frac{\md}{\bfnd(\k)} \right) \right)^{\Cdot[2]} = \e^{2\nu} \left( \scA(\k) - \md  \scAb(\k) \right) \nonumber
    \end{equation}
    here, we have allowed for a relative scaling by $\md$, which is a consistent rescaling within a scheme: the coefficients and the couplings need to be simultaneously scaled by $\md$ and $1/\md$, respectively, which leaves the $\b$-function equations invariant. Defining, $ c(\k) = \dfrac{1}{\fnd(\k)} - \dfrac{\md}{\bfnd(\k)}$, which can be viewed as an expansion by itself, $c(\k) = c_0 + \k^2 c_2 + \k^4 c_4 + \cdots$, we can solve for $c(\k)$,\footnote{this expansion is motivated by RHS of the equation, and has some non-trivial implication. Since the relation between $\bfnd$ and $\fnd$ doesn't depend explicitly on $t$, this can be directly understood as a diffeomorphism in the space of couplings.}
    \begin{align}\label{eq:c-sol-integral}
	e^{2\nu t} c(e^t k) - \lim_{t\rightarrow-\infty} \left( e^{2\nu t} c(e^t k) \right) &= \int_{-\infty}^t dt \ e^{2\nu t} \left[ \scA(e^t k) - \md \scAb(e^t k) \right]
    \end{align}
    here we have used, $\k= \e k$ and the redefinition $\e=e^t$. Solving the above equation \eqref{eq:c-sol-integral} term by term as a series in $\k$, we get,
    \begin{align}\label{eq:c-sol}
	&& c(e^t k) = \sum_{j=0} \left[ \left(e^{2t}\right)^j k^{2j} c_j \right] &= \sum_{j=0} \left(e^{2t}\right)^j k^{2j} \ \frac{\scA_j - \md \scAb_j}{2\nu + 2 j} && \nonumber \\
	&& \hspace{1cm} c_i &= \frac{\scA_i - \md \scAb_i}{2\nu + 2 i}, \hspace{1cm} i \ge 0 &&
    \end{align}
    The relation $ c(\k) = \dfrac{1}{\fnd(\k)} - \dfrac{\md}{\bfnd(\k)}$ gives us a transformation in the coupling-space which relates the two RG-schemes at an arbitrary cut-off.

\section{Discussions}\label{sec:discussions}
In this paper we have determined all possible boundary conditions for
a single bulk scalar field in AdS/CFT. The principle is that these
boundary conditions can be regarded as wavefunctionals whose
$z$-dependence is determined by a radial Schr\"{o}dinger equation. We found that the original GKPW prescription, coupled with the Solodukhin
counterterms and applied to a finite radial cut-off $z=\e_0$, corresponds to a wavefunctional which cannot be obtained by the evolution of the known GKPW $\d$-function boundary condition at $z=0$. In addition,  it contains some spurious double trace deformations. We
found a precise field theory correspondence for all allowed boundary conditions and
found two specific wavefunctions (boundary conditions) $\Psi_1^0$ and
$\Psi_2^0$ (\autoref{eq:wavefunctional-correct-st} and \ref{eq:wavefunctional-correct-aq}) which represent the pure CFTs (respectively, IR and UV CFT,
corresponding to standard and alternative quantizations). Using this
insight, we isolate the real double trace deformations from spurious
ones and find that the holographic beta-functions can be matched to
the ones computed from field theory. We gave a geometric interpretation of the specific
wavefunctionals in terms of a specific form of non-locality of the
boundary `points' in Witten diagrams.

As mentioned above, we have discussed the field theory equivalent of the above boundary wavefunctional in terms of
properties of the generating functional $Z[J]$. In field theory, it is in
principle possible, though difficult in practice,
to reproduce the continuum result (power law scaling) at a finite
cut-off scale, in terms of effective Wilsonian
vertices plus a $J^2$ term in $\log Z[J]$\footnote{We thank
Shiraz Minwalla for discussion on this point}. However, holography gives
such an `RG scheme' in a rather straightforward fashion.

In this paper, we considered a probe approximation; it was sufficient
for our purposes to consider a quadratic bulk scalar action. We expect that for an interacting bulk action, with possibly multiple
fields, it should again be possible to discover boundary wavefunctionals
defining AdS/CFT at a finite cut-off, such that the pure CFT
correlators are reproduced at a finite
cut-off. The argument for the
existence of such boundary conditions follows from the abstract
argument, presented above for existence of such RG schemes in field
theory. It is also
interesting to speculate what the appropriate AdS/CFT prescription 
at a finite cut-off is for scalar fields in a black hole background.  As against the pure
AdS background, we now expect the correct wavefunctionals $\Psi_1^0$ (and
$\Psi_2^0$, when the alternative quantization exists), to have
a specific dependence on the new scale provided by the black hole
horizon radius. We hope to come back to these issues shortly.

\section*{Acknowledgements}
We thank Shamik Banerjee, Avinash Dhar, Sachin Jain,
Nilay Kundu, Sung-Sik Lee, Shiraz Minwalla,  Mukund
Rangamani, Arnab Rudra, Ronak Soni, Sandip Trivedi and David Tong for enlightening
discussions, and Shiraz Minwalla for insightful comments on an early
draft. 

This work was supported in part by Infosys Endowment for the study of the Quantum Structure of Space Time.

\pagebreak
\begin{appendices}
\appendix

\section{\label{sec:notations}Notations}

We will generally use the notation $f_n$ for the double-trace
couplings introduced in \eq{all-double-trace}, and $\bar{f}_n$ for
their dimensionless counterparts. In addition, depending on the
context we will denote these couplings by the following specialized
notations:

\begin{center}
\begin{tabular}{l|l|l}
\toprule
Field Theory&
Dimensionful: $\fd$&
Dimensionless: $\fnd$\\
\midrule
Bulk&Dimensionful: $\bfd$& 
Dimensionless: $\bfnd$\\
\bottomrule
\end{tabular}
\end{center}

\section{\label{app:results}Some mathematical results}

  \subsection*{Integrals}
  Integrals of the following type appear in the calculation of $\b$-functions,
      \begin{align}\label{eq:app-integral}
       &\int\limits_{a'>|w|>a} d^dw \, \dfrac{w^{\mu_1} w^{\mu_2} \cdots w^{\mu_{2n}}}{|w|^{p}} = \int\limits_{a'>w>a}  dw \dfrac{1}{|w|^{p-d+1-2n}} \int d\Omega_{d-1} \hat{w}^{\mu_1} \hat{w}^{\mu_2} \cdots \hat{w}^{\mu_{2n}} \nonumber \\
       &= \frac{1}{d+2n-p} \left( a'^{d+2n-p} - a^{d+2n-p} \right) \nonumber \\
       &\times \left( \dfrac{2^{1-2n} \pi^{d/2}}{\Gamma\left(\dfrac{d}{2}+n\right) \Gamma(n+1)} \right) \sum_{\PermP\in S_{2n}} \Big( \d^{\mu_{\PermP(1)} \mu_{\PermP(2)}} \; \d^{\mu_{\PermP(3)} \mu_{\PermP(4)}} \cdots \d^{\mu_{\PermP(2n-1)} \mu_{\PermP(2n)}} \Big)
      \end{align}
      here, $\PermP$ runs over all permutations of $2n$ numbers, and hence a lot of terms in the parenthesis in the last line are equivalent. The pre-factor has been accordingly calculated to account for these redundancies. This convention is useful because contractions of the $(2n)!$ different permutations of Kronecker-$\d$ above with $\del_{\mu_1}\ldots\del_{\mu_{2n}}$ generates $(2n)! (\del^2)^n$, and the $(2n)!$ here exactly cancels with $1/(2n)!$ coming from the Taylor series. We have used the following short-hand notation for the pre-factor in the paper,
      \begin{align}
	  \a_n = \left( \dfrac{2^{1-2n} \pi^{d/2}}{\Gamma\left(\dfrac{d}{2}+n\right) \Gamma(n+1)} \right)
      \end{align}
      This factor also obeys an identity,
      \begin{align}\label{eq:alpha-identity}
	  \dfrac{\a_n}{\a_{n+1}} = 2 (n+1) (d+2n)
      \end{align}
      which is useful in simplifying the coefficients of the $\b$-functions.

  \subsection*{Variation of derivatives of propagators}
      Coefficients of all the terms in the $\b$-function equations are of the form (see \eqref{eq:quadratic-term-in-beta}), $$ \left( \int d\rho \ \rho^{d-1+2k} \ \del_{a'}\left[(a')^{-2n} (\del_\rho^2)^n G_{a'}(a' \rho)\right] \right) $$
      In this appendix, we list first few expressions for  $(\del^2)^n G_{a'}(w)$ and $ \del_{a'}\left[ (\del^2)^n G_{a'}(w) \right] $, in terms of various derivatives of the regulating function, $\mK^{(n)}$.
      
\LTcapwidth=6in
      \begin{longtable}[H]{c|c}
	      \toprule[1.7pt]
	      \textbf{n}&\boldmath$ (\del^2)^n G_{a'}(w) $\\
	      \midrule[1.5pt]
	      0&$ \dfrac{\mK(w/a) }{ w^{2 \Delta}} $\\
	      \midrule
	      1&$ \dfrac{ \mK''(w/a)}{a^2 \ w^{2 \Delta }}- (4 \Delta - d +1) \dfrac{ \mK'(w/a)}{a \ w^{2 \Delta +1}}+2 \Delta  (2 \Delta - d +2) \dfrac{\mK(w/a) }{ w^{2 (\Delta +1)}} $\\
	      \midrule
	      2&$ \dfrac{\mK^{(4)}(w/a)}{a^4 \ w^{2 \Delta }} - 2 (4 \Delta - d +1) \dfrac{ \mK^{(3)}(w/a)}{a^3 \  w^{2 \Delta +1}} + \Big(d^2-4 d (3 \Delta +1)+3 \left(8 \Delta ^2+8 \Delta +1\right)\Big) \times $ \\
	      &$\hspace{2cm} \dfrac{ \mK''(w/a)}{a^2 \ w^{2 (\Delta +1)}} - (4 \Delta -d +3) (4 \Delta  (2 \Delta +3) - 4 d \Delta - d + 1) \dfrac{ \mK'(w/a)}{a \ w^{2 \Delta +3}} $\\
	      &$\hspace{3cm} +4 \Delta  (\Delta +1) (2 \Delta - d +4) (2 \Delta -d +2) \dfrac{\mK(w/a)}{ w^{2 (\Delta +2)}} $\\
	       \midrule
	      $\vdots$&$\vdots$\\
	      \bottomrule[2pt]
 	\caption{List of various powers of Laplacian acting on propagator $G_{a'}(w)$, which are needed in the computation of $\b$-functions.}
 	\label{tab:List-Derivatives}
      \end{longtable}
\vspace{-0.35cm}
      \begin{longtable}[H]{c|c}
	      \toprule[1.75pt]
	      \textbf{n}&\boldmath$ \del_{a'}\left[ (\del^2)^n G_{a'}(w) \right] $\\
	      \midrule[1.5pt]
	      0&$ - \dfrac{ \mK'(w/a)}{a^2 \  w^{2 \Delta -1 }} $\\
	      \midrule
	      1&$ -\dfrac{\mK^{(3)}(w/a)}{a^4 \ w^{2 \Delta - 1 } } + (4 \Delta - d - 1) \dfrac{ \mK''(w/a)}{a^3 \ w^{-2 \Delta } } - \left(4 \Delta ^2-2 d \Delta +d-1\right)  \dfrac{ \mK'(w/a)}{a^2 \ w^{2 \Delta +1}} $\\
	      \midrule
	      2&$ -\dfrac{ \mK^{(5)}\left(w/a\right)}{a^6 \  w^{2 \Delta -1}} + 2  (4 \Delta - d -1) \dfrac{ \mK^{(4)}\left(w/a\right)}{a^5 \ w^{2 \Delta } } - \left(24 \Delta ^2-12 d \Delta +d (d+2)-3\right) \dfrac{ \mK^{(3)}\left(w/a\right)}{a^4 \ w^{2 \Delta +1}}$\\
	      &$\hspace{3cm}+ (4 \Delta - d +1)  \left(8 \Delta ^2-4 d \Delta +4 \Delta + d -3\right)  \dfrac{ \mK''\left(w/a\right)}{a^3 \ w^{2 (\Delta +1)}} $ \\
	      & $\hspace{3.5cm} - \left(4 \Delta ^2-1\right)  (2 \Delta -  d + 3) (2 \Delta - d +1) \dfrac{ \mK'\left(w/a\right)}{a^2 \ w^{-2 \Delta -3}} $\\
	       \midrule
	      $\vdots$&$\vdots$\\
	      \bottomrule[2pt]
       	\caption{List of variation of $(\del^2)^n G_{a'}(w)$ with respect to $a'$.}
       	\label{tab:list-variation}
      \end{longtable}

      These expressions listed in \autoref{tab:list-variation} are part of the integrands that appear in \eqref{eq:quadratic-term-in-beta}. For a general coefficient, we use following notation for these integrals \eqref{eq:short-hands},
      \begin{align}
	    \Gcoff^{\mK^{(n)}}_{\D} &= \int d\rho \ \rho^{d-2\D} \ \mK^{(n)}(\rho)
      \end{align}
      and corresponding values from \autoref{tab:list-variation} have been used to exactly compute the coefficients in \eqref{eq:General-Beta}, \eq{eq:Beta-FT-final} and \autoref{tab:Couplings} for the choices $\mK(\rho) = \Theta(\rho-1)$ and $\mK(\rho) = \text{regulated-}\Theta(\rho-1)$ for the regulating function.

  \section{Holographic Wilsonian Renormalization: Explicit solution}\label{app:Eff-Act}
  In this appendix we compute the $\b$-functions using Holographic Wilsonian RG techniques. Calculations are based on the work that appears in \cite{Heemskerk:2010hk,Faulkner:2010jy,Elander:2011vh} with modifications called for by introducing finite cut-off as discussed in \autoref{sec:implications}.

As explained in \autoref{sec:hol-RG} we separate the bulk degrees of freedom into UV and IR degrees of freedom and integrate out the near boundary (UV) degrees of freedom, as we change the radial cut-off surface from $z=\e_0$ to $z=\e$. In the process we generate a modified wavefunctional $\Psi[\phi_0;\e] = Z_{UV}$ at the new boundary $z=\e$, whose coefficients contain that information about the couplings of double-trace operators in the field theory at the new cut-off as given by \eqref{eq:dbl-trace-bulk-st} and \eqref{eq:dbl-trace-bulk-aq}.

The bulk evolution equation in radial direction can be determined by computing the \emph{radial Hamiltonian}.
  \begin{align}
    \mathcal{H} &=  \frac{1}{2} \left( \frac{\pi^2}{z^{1-d}} + \frac{z^{-1-d}}{2} \left( \del^\mu \phi \del_\mu \phi + m^2 \phi^2 \right) \right)
  \end{align}
  in operator language, the evolution Hamiltonian in the radial direction is,
  \begin{equation}
    \hat{H} = \int d^dx \ \hat{\mathcal{H}} = \frac{1}{2} \left( \int d^d k \ \frac{1}{z^{1-d}} \hat{\Pi}_k \hat{\Pi}_{-k} + z^{-1-d} \left(z^2 k^2+m^2 \right) \hat{\phi}_k \hat{\phi}_{-k} \right)
  \end{equation}
  here $\hat{\Pi} \equiv i \frac{\d}{\d \phi}$ in the `field basis', where $\hat{\phi}(x)\left|\phi\rangle\right. = \phi(x)\left|\phi\rangle\right.$. The radial Schr\"{o}dinger equation for the radial wavefunctional $Z_{UV}$ is given by \eqref{eqn:evolution}, \footnote{In the particular case of quadratic bulk action, the case we are demonstrating here, the Schr\"{o}dinger equation and the semi-classical Hamiltonian-Jacobi equations are equivalent.}
  \begin{align}
    &-\del_\e Z_{UV} = \hat{H} Z_{UV} \nonumber
  \end{align}
  Since we are working with a quadratic theory and the boundary wavefunctional at $z=\e_0$ is also quadratic, the wavefunctional generated at any other cut-off $z=\e$, $Z_{UV} = \Psi[\phi_\e;\e]$ is also quadratic. So let us consider a general form of the wavefunctional,
  \begin{align}
  Z_{UV} = \exp \left[ -\half \int d^dk \sqrt{\g} \left( A(k\e;k\e_0) \phi^\e_{k} \phi^\e_{-k} + 2 \e^{d-\D} B(k\e;k\e_0) J^0_{k} \phi^\e_{-k}\right.\right.\nonumber\\ 
  \left.\left.+ \e^{2(d-\D)} C(k\e;k\e_0) J^0_{k} J^0_{-k} \right)\right]
  \end{align}
  to keep the calculation more general, we don't specify $\D$ here. In subsequent computations $\D=\D_+$ for standard quantization and $\D=\D_-$ for alternative quantization. We now derive the general evolution equations for the coefficients $ A(k,\e,\e_0),  B(k,\e,\e_0),$ $C(k,\e,\e_0)$. The exact form of these coefficients can be obtained by starting with the appropriate wavefunctionals \eqref{eq:dbl-trace-bulk-st} or \eqref{eq:dbl-trace-bulk-aq} at $z=\e_0$ but since the evolution equation doesn't depend on the initial wavefunctional it is not required here.  When substituted in the radial Schr\"{o}dinger equation we get,
  \begin{align}
    -\del_\e Z_{UV} &= \Bigg( \half \int d^dk \left[ \del_\e \left( \e^{-d} A(k\e;k\e_0) \right) \phi^\e_{k} \phi^\e_{-k}  + 2 \del_\e \left( \e^{-\D} B(k\e;k\e_0) \right) \phi^\e_{k} J^0_{-k} \right. \nonumber \\
    & \hspace{5cm} \left. + \del_\e \left( \e^{d-2\D} C(k\e;k\e_0) J^0_{k} J^0_{-k} \right)\right] \Bigg) \times Z_{UV} \nonumber \\
    \hat{H} Z_{UV} &= \Bigg( \half \int d^dk \e^{-d-1} \Big[  \left( \e^2k^2+m^2-A^2(k\e;k\e_0) \right) \phi^\e_{k} \phi^\e_{-k} \nonumber \\
    & - 2 A(k\e;k\e_0) B(k\e;k\e_0) \e^{d-\D} \phi^\e_{k} J^0_{-k} - \e^{2(d-\D)} B^2(k\e;k\e_0) J^0_{k} J^0_{-k} \Big] + \cdots \Bigg) \times Z_{UV}
  \end{align}
  the terms in the ellipsis in the above equation are not important and don't arise when we keep track of the overall normalisation of $Z_{UV}$. $J_0$ above is the source for the operator $\mO$ at $z=\e_0$. This implies following evolution equations for the coefficients,
  \begin{subequations}\label{eq:ABC-beta}
    \begin{equation}
	\e\del_\e A = - A^2 + dA+ (\e^2 k^2 + m^2)
    \end{equation}
    \begin{equation}
     	\e\del_\e B = \D \; B - A \; B
    \end{equation}
    \begin{equation}
	\e\del_\e C = (2\D-d) \; C - B^2
    \end{equation}
  \end{subequations}
  the field theory double-trace couplings are related to $A(k,\e,\e_0)$ by \eqref{eq:dbl-trace-bulk-st} and \eqref{eq:dbl-trace-bulk-aq} (recall, $\bfnd$ denotes dimensionless coupling),
  \begin{subequations}
    \begin{equation}
	\bfnd^{ST}(k^2\e^2) = \frac{ \Big( A(k\e;k\e_0) -\DCT(k\e) \Big) \ \afxd-1}{ { B^*_{ST} }^{2} \bfnd^{*}_{ST}{}^{2}  \Big( A(k\e;k\e_0)-\DCT(k \e) \Big) }
    \end{equation}
while, with the inclusion of the counter-term,
    \begin{equation}
	\bfnd^{ST}(k^2\e^2) = \frac{ \Big( A(k\e;k\e_0) -\DCT(k\e) \Big) \ \afxd-1}{ \left( { B^*_{ST} }^{2} \bfnd^{*}_{ST}{}^{2} + \afxd \cdot \d C \right)  \Big( A(k\e;k\e_0)-\DCT(k \e) \Big) - \d C }
    \end{equation}
    \begin{equation}
	\bfnd^{AQ}(k^2\e^2) = \frac{A(k\e;k\e_0)-\DCT(k\e)}{ C^*_{AQ} \left( A(k\e;k\e_0) - \DCT(k\e;k\e_0) \right) + { B^*_{AQ} }^{2} }
    \end{equation}
and with the inclusion of the counter-terms,
        \begin{equation}
	\bfnd^{AQ}(k^2\e^2) = \frac{A(k\e;k\e_0)-\DCT(k\e)}{ \left( C^*_{AQ} + \d C \right) \left( A(k\e;k\e_0) - \DCT(k\e;k\e_0) \right) + { B^*_{AQ} }^{2} }
    \end{equation}
  \end{subequations}
  Equations \eqref{eq:ABC-beta} can be used to compute the $\b$-function equations for these couplings,\\
  For standard quantization (see \eqref{eq:st-bulk-beta-dC}),
    \begin{align}
	\e \del_\e \bfnd =&\ \bfnd^2 \times \Bigg( {B^*_{ST}}^2 \ {\afxd}{}^2 \left( k^2 \e^2+ m^2 + \DCT (d-\DCT)-\e \del_\e\DCT \right) - \e \del_\e \d C  \nonumber \\
	& \hspace{0.25cm} + \frac{(\d C)^2}{{B^*_{ST}}^2 \afxd{}^2}  \left( \e \del_\e \bfnd^*-1 + \afxd (d-2 \DCT)+\afxd{}^2 \left( k^2 \e^2 + m^2 + \DCT (d-\DCT)-\e \del_\e \DCT \right) \right) \nonumber \\
	& \hspace{-0.28cm} - \d C \left( -\frac{2 \e \del_\e B^*_{ST}}{B^*_{ST}} - \frac{2 \ \e \del_\e{\afxd}}{\afxd} - 2 {\afxd} \left( k^2 \e^2 + m^2 + \DCT (d-\DCT)-\e \del_\e \DCT \right) -d+2 \DCT \right) \Bigg) \nonumber \\
	    & \hspace{-6.5pt}+  \bfnd \Bigg(-\frac{2 \ \e \del_\e{B^*_{ST}}}{{B^*_{ST}}} - \frac{2 \ \e \del_\e{\afxd}}{\afxd} -2 {\afxd} \left( k^2 \e^2+ m^2 + \DCT (d-\DCT)-\e \del_\e\DCT\right) -d+2 \DCT(k \e) \nonumber \\
	& \hspace{-0.28cm}  - 2 \dfrac{\d C}{{B^*_{ST}}^2 \afxd{}^2} \left( \e \del_\e \bfnd^*-1 + \afxd (d-2 \DCT)+\afxd{}^2 \left( k^2 \e^2 + m^2 + \DCT (d-\DCT)-\e \del_\e \DCT \right) \right) \Bigg) \nonumber \\
	&\hspace{1.5cm} + \ \frac{ \e \del_\e{\afxd} + \afxd \ (d-2 \DCT)+\afxd{}^2 \left( k^2 \e^2+ m^2 + \DCT (d-\DCT)-\e \del_\e\DCT\right) -1}{{B^*_{ST}}^2 \ {\afxd}{}^2} \nonumber
    \end{align}

    And for alternative quantization \eqref{eq:aq-bulk-beta-dC},
	\begin{align}
		\e \del_\e \bfnd =& \frac{1}{{B^*_{AQ}}^2} \Bigg[\bfnd^2 \bigg(2 {B^*_{AQ}} \ \e \del_\e {B^*_{AQ}} \left( {C^*_{AQ}} + \d C \right) -{B^*_{AQ}}^2 \left(\e \del_\e {C^*_{AQ}} + \e \del_\e \d C+ \left({C^*_{AQ}} + \d C \right) (d-2 \DCT)\right) \nonumber \\
		&\hspace{4cm}  - {B^*_{AQ}}^4 + \left( {C^*_{AQ}} + \d C \right)^2 \left(\DCT (d-\DCT)-\e \del_\e \DCT+k^2 \e^2+m^2\right) \bigg) \nonumber \\[5pt]
		&\hspace{-0.85cm}+\bfnd \left(-2 {B^*_{AQ}} \ \e \del_\e {B^*_{AQ}} +{B^*_{AQ}}^2 (d-2 \DCT)-2 \left( {C^*_{AQ}} + \d C \right) \left(\DCT (d-\DCT)-\e \del_\e \DCT+k^2 \e^2+m^2\right)\right) \nonumber \\[5pt]
		&\hspace{3cm} + \DCT (d-\DCT)-\e \del_\e \DCT+k^2 \e^2+m^2 \Bigg] \nonumber
	\end{align}

    In the above equations, we have suppressed the functional dependence of $\DCT(k\e), {B^*_{ST}}(k\e)$ and $ \afxd(k\e)$ to avoid clutter. Although the above equations look horrendous, when resolved in components of the coupling $\bfnd = \bfnd_0 + \bfnd_1 (k\e)^2 + \bfnd_2 (k\e)^4 + \cdots$, and on substituting the values of $\DCT(k\e), {B^*_{ST}}(k\e)$ and $\afxd(k\e)$ given by \eqref{contact},  \eq{eq:ABC-series-sol-st},  \eq{eq:ABC-series-sol-aq}, the $\b$-functions for individual couplings become quite simple,\\
    \textbf{Standard Quantization:}
     \begin{equation}\begin{aligned}\label{eq:st-bulk-indiv-beta-2}
	& \dot{\bfnd}_0 =  - 2 \nu \ \bfnd_0 + 2 \nu \ c_0 \ \bfnd_0^2  \\
	& \dot{\bfnd}_1 =  - (2 \nu+2) \ \bfnd_1 - 2 (1-\nu) \ c_1 \ \bfnd_0^2  + 4 \nu \ c_0 \ \bfnd_0 \bfnd_1   \\
	& \dot{\bfnd}_2 =  -  (2 \nu+4) \ \bfnd_2 - 2 (2-\nu) \ c_2 \ \bfnd_0^2 - 4 (1-\nu) \ c_1 \  \bfnd_0 \bfnd_1 + 4 \nu \ c_0 \ \bfnd_0  \bfnd_2 + 2 \nu \ c_0 \ \bfnd_1^2 \\
	& \hspace{1cm} \vdots
    \end{aligned}\end{equation}

\textbf{Alternative Quantization:}
    \begin{equation}\begin{aligned}
    \label{eq:aq-bulk-indiv-beta-2}
	& \dot{\bfnd}_0 = 2 \nu \bfnd_0  - 2 \nu \ c_0 \ \bfnd_0^2  \\
	& \dot{\bfnd}_1 = (2 \nu-2) \bfnd_1 -2 (1+\nu) \ c_1 \ \bfnd_0^2 - 4\nu \ c_0 \ \bfnd_0 \bfnd_1  \\
	& \dot{\bfnd}_2 = (2 \nu-4) \bfnd_2 -  2(2+\nu) \ c_2 \ \bfnd_0^2 - 4 (1+\nu) \ c_1 \ \bfnd_0 \bfnd_1 - 4 \nu \ c_0 \ \bfnd_2 \bfnd_0 - 2 \nu \ c_0 \ \bfnd_1^2  \\
	& \hspace{1cm}\vdots \\
    \end{aligned}\end{equation}
    The fixed point values for the coupling constants given by solving the stationary points of the above equations are (for both standard and alternative quantization),
    \begin{align}
	\text{Trivial Fixed Point:} & \ \bfnd_i = 0 \ \forall \ \{ i \in \mathbb{Z}^+ \cup 0 \} \nonumber \\
	\text{Non-Trivial Fixed Point:} & \ \bfnd_0\to \frac{1}{c_0} ,\ \bfnd_1\to -\frac{c_1}{c_0^2},\ \bfnd_2\to \frac{c_1^2 - c_0 c_2}{c_0^3} \ldots \label{eq:Fixed-pt}
    \end{align}
    
    It might look strange that the fixed point for both standard and alternative quantization in \eqref{eq:Fixed-pt} is the same. This happens because the counter-terms, $\d C$ in one theory aren't the same as those in the other theory. Here we have only used them as a notational device and so they should not be confused to be equivalent. We discuss the relation between the non-trivial fixed points of one theory with the trivial fixed point of the other theory in the next subsection.

\subsection{Relation between Standard and Alternative Quantizations\label{app-sec:Comments-Standard}}
We had remarked in \autoref{sec:alternate} how the undeformed alternative and standard quantized theories are Legendre transform of each other. This relationship doesn't hold exactly anymore for the regulated theories given by the inclusion of \eqref{eq:Sextra-st} and \eqref{eq:Sextra-aq}. 
However, as one would expect, the UV fixed point of the regulated standard quantized theory is the alternative theory and vice versa. In the following discussion we show this relationship explicitly.

From \eq{eq:Fixed-pt} we see that the non-trivial fixed point corresponds to couplings $\bfnd(k^2 \e^2) = \dfrac{1}{\d C(k\e)} $. So the correlators at the non-trivial fixed points are given by, \eq{eq:corr-def-st} and \eq{eq:corr-def-aq},
    \begin{subequations}\begin{align}
    	\lan \mO(k) \mO(-k) \ran^{fp}_{+} = \frac{k^{2\nu} \ \dfrac{2^{1-2\nu }  \Gamma (1-\nu )}{\Gamma (\nu )} + \e^{-2\nu} \ \d C_{ST}(k\e)}{2 + \dfrac{(k\e)^{2\nu}}{\d C_{ST}(k\e)} \ \dfrac{2^{1-2\nu }  \Gamma (1-\nu )}{\Gamma (\nu )} }\label{eq:nft-corr-st}\\[15pt]
	\lan \mO(k) \mO(-k) \ran^{fp}_{-} = \frac{-k^{-2\nu} \ \dfrac{2^{2\nu -1}  \Gamma (\nu )}{\Gamma (1-\nu )} + \e^{2\nu} \ \d C_{AQ}(k\e)}{2-\dfrac{(k\e)^{-2\nu}}{\d C_{AQ}(k\e)} \ \dfrac{2^{2\nu-1 }  \Gamma (\nu )}{\Gamma ( 1-\nu )} }\label{eq:nft-corr-aq}
    \end{align}\end{subequations}
here, the superscript $fp$ signifies that we are computing the correlator at the non-trivial fixed point of the theory. The flow towards UV starting from the standard quantization is defined by taking the limit $k\e\to\infty$ in \eq{eq:nft-corr-st}. In this limit the correlation function becomes,
\begin{align}\label{eq:nft-corr-st-uv}
	\left. \lan \mO(k) \mO(-k) \ran^{fp}_{+}\right|_{k\e\to\infty} = \left( \e^{-2\nu} \d C_{ST} \right)^2 \left[ \dfrac{\e^{2\nu}}{\d C_{ST}} - k^{-2\nu} \ \dfrac{2^{2\nu -1}  \Gamma (\nu )}{\Gamma (1-\nu )} \right]
\end{align}
which is the same as the correlator of the regulated alternative theory if we identify $\d C_{AQ} = 1/\d C_{ST}$, upto some overall multiplicative wavefunctional renormalization, $\mO_-(k) = (\e^{-2\nu} \d C_{ST})^{-1}  \cdot \mO_+(k) = \bfd_+^* \mO_+(k)$.\footnote{This wavefunctional renormalization is well known in the literature and provides for the correct scaling dimension of the operators at the non-trivial fixed point.\\
Also, for clarification of notation, $\bfd^*_\pm$ are the non-trivial fixed points for the standard and alternative theories.\\
$\mO_+$ and $\mO_-$ are the operators dual to the bulk field $\phi$ at the standard and alternative fixed points respectively.}
Similarly for the flow towards IR fixed point from the alternative fixed point, we take the IR limit, $k\e\to0$ in \eq{eq:nft-corr-aq},
\begin{align}\label{eq:nft-corr-aq-ir}
	\left. \lan \mO(k) \mO(-k) \ran^{fp}_{-}\right|_{k\e\to0} = \left(\e^{2\nu} \d C_{AQ}\right)^2 \left[ \frac{\e^{-2\nu}}{\d C_{AQ}} + k^{2\nu} \ \dfrac{2^{1-2\nu} \Gamma(1-\nu)}{\Gamma(\nu)} \right]
\end{align}
which again is the same as the correlator of the regulated standard theory with the identification $\d C_{ST} = 1/\d C_{AQ}$, and $\mO_+(k) = (\e^{2\nu} \d C_{AQ})^{-1}  \cdot \mO_-(k) = \bfd_-^* \mO_-(k)$. Thus clearly, the standard quantized theory and alternative quantized theory are connected to each other with RG flow as IR and UV fixed points.

All the results discussed here are parallel to the field theory calculations that were presented in \autoref{sub-sec:ft-corr-mom-sp}.

\section[\texorpdfstring{Large $N$ limit of $O(N)$ Wilson-Fisher model}{Large N limit of O(N) Wilson-Fisher model}]{Large $N$ limit of $O(N)$ Wilson-Fisher model 
\label{app:wilson-fisher}}

Let us consider the following Euclidean action in $d=4- \ep$
dimensions (see, e.g. \cite{Moshe:2003xn})
\[
S = \int d^d x \left\{  \f12 \left(\del_\mu \phi_i\right)^2 +\f12 m_0^2
\mO(x) + \f1{4!} \f{g_0}{N} \L^{\e}   \mO(x)^2 \right\}, \quad \mO(x)=  
\phi_i\phi_i(x)
\]
The phase diagram and fixed points of this model are shown in Fig.
\ref{fig:moshe}. The model possesses a critical surface (where the
correlation length diverges) given by
\[
m_0^2= - g_0 \f16  \L^{\e} \Omega_d(0), \;
 \Omega_d(0)\equiv \f1{(2\pi)^d} \int^\L \f{d^dk}{k^2}  \propto \L^{d-2}
\]
\begin{figure}[!th]
    \centering
    \includegraphics[scale=0.15,keepaspectratio=true]{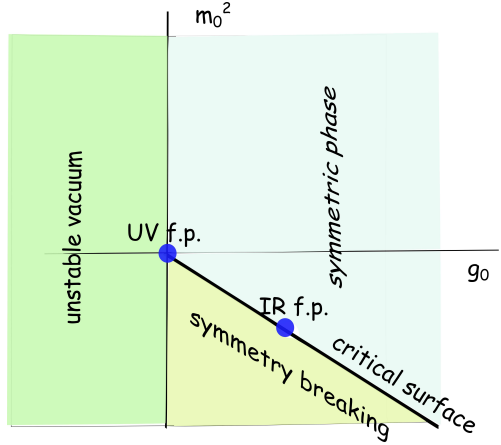}
    \caption{Large-$N$ Wilson-Fisher: fixed points and phase diagram.}
    \label{fig:moshe}
\end{figure} 
The $\b$-function is given by 
\[
\L\del_\L g_0= \b(g_0)= - \e g_0 + \f{N+8}{N} \f{g_0^2}{48 \pi^2}  + O(g_0^3)
\]
which shows a UV fixed point at $g_0=0$ and an IR fixed point at 
\[
g_0^* =  \e  \f{48 \pi^2 N}{N+8}  + O(\e^2)
\] 
The two-point function of $\mO(x)$ can be obtained in
the large $N$ limit by saddle point methods, and is given by (see
Sections 2.3 and 2.4 of \cite{Moshe:2003xn}, especially Eqs.  (2.57)
and (2.59))
\begin{align}
& \lan \mO(p) \mO(q) \ran = G(p) \d(p+q),
\;  G(p) = - \L^{-\e} \f{\f{12}{g_0}}{1 + \L^\e \f{g_0}6 B_\L(p)}
\nonumber\\
& B_\L(p) = \int^\L \f{d^dk}{k^2 (k-p)^2} =  p^{-\e}\left(b_0 +
b_1 (p/\L)^2 + \cdots\right) + \L^{-\e}\left(a_0 + a_1  (p/\L)^2 + 
\cdots\right)
\label{bubble} 
\end{align}
where $b, a$ are some constants. 

\paragraph{The IR behaviour:} IR limit is given by $p/\L \to 0$, 
\begin{align}
G_{IR}(p) &= -\frac{72 \L^{-2\e}}{g_0^2 Z^2} p^{\e} \ \left[ 1 + \left( \frac{p}{\L}\right)^{\e}  \left(\d C + \frac{6}{g_0 Z^2} \right) \right]^{-1} \nonumber \\
\underrightarrow{{\scriptstyle{p/\L\to\infty}}} &\ -\frac{72 \L^{-2\e}}{g_0^2 Z^2} p^{\e}
\label{wilson-fisher-2pt-ir}
\end{align} 
where, we have used the notation, $Z^2 = \left(b_0 + b_1 (p/\L)^2 + \cdots\right), \ Z^2 \cdot \d C = \left(a_0 + a_1  (p/\L)^2 + \cdots\right)$. The renormalized IR operators are given by $\mO_{IR} = \left(\frac{g_0  \; \L^\e}{12}  \right)^2 \mO_{UV}$, which is well known for the Wilson-Fisher fixed point.\footnote{Note that there is a slight difference in the correlator here compared to \autoref{sub-sec:ft-corr-mom-sp} because the correlator in \eq{bubble} is not of the form $\frac{G}{1+\fd G}$, and the conventions in \cite{Moshe:2003xn} are such that the IR correlator appears without the contact-terms.}

\paragraph{The UV behaviour:}
In the limit $p/\L \to \infty$, we get\footnote{Note that the normalization of the two-point function differs from the main text,
due to a different normalization of the operator $\mO(x)$. We can identify correctly normalized UV operator as, $\mO_{UV} = \sqrt{2} \; Z \; \mO$}
\begin{align}
G_{UV}(p) &= -\frac{\frac{12\L^{-\e}}{g_0}}{1+\frac{g_0}{6}Z^2 \cdot \d C} \left[ 1 + \frac{\frac{g_0}{6} Z^2}{1+\frac{g_0}{6}Z^2 \cdot \d C} \left( \frac{\L}{p} \right)^\e \right]^{-1} \nonumber \\[5pt]
\underrightarrow{{\scriptstyle{p/\L\to\infty}}} &\ -\frac{\frac{12\L^{-\e}}{g_0}}{1+\frac{g_0}{6}Z^2 \cdot \d C} + \frac{ 2 Z^2 }{ \left( 1+\frac{g_0}{6}Z^2 \cdot \d C \right)^2} \ p^{-\e}
\label{wilson-fisher-2pt-uv}
\end{align} 
which again agrees with the general analysis presented in \autoref{sub-sec:ft-corr-mom-sp}, upto some normalization and contact terms which can be attributed different regulation used in \cite{Moshe:2003xn}.

\section[\texorpdfstring{Large $N$, Probe approximation and Hamilton-Jacobi}{Large N, Probe approximation and Hamilton-Jacobi}]{Large $N$, Probe approximation and Hamilton-Jacobi
\label{app:probe-HJ}} 

\paragraph{Probe approximation:} 

Let us consider a free massive scalar field described by
\eq{bulk-action} but coupled to a perturbed metric of the form
$g_{MN}= \bg_{MN} + \sqrt{\k}\ h_{MN}$ where $\bg_{MN}$ is now the AdS
metric \eq{ads-metric}. In this case the bulk action is of the
schematic form (where we focus on the $\k$-dependence)
\begin{align}
& S \sim S_b + S_{grav} + S_{int}, 
\nonumber\\
& S_b \sim \int  (\del
\phi)^2 + m^2 \phi^2, \quad S_{grav} \sim \int (\del h)^2,
\quad S_{int} \sim \int \sqrt{\k} (h \del\phi\del\phi + h \del h \del
h) + \k\ h h \del h \del h
\label{bulk-action-pert}
\end{align}
The bulk partition function, computed from the above, clearly matches
(in large $N$ counting) a field theory partition of the form $\lan
\exp[\int \phi_0(x) \mO(x)]$ where the connected two-point function is
normalized as $\lan \mO \mO \ran \sim O(1)$. The connected 3-point
function $\lan \mO \tilde T \tilde T $ (where $\tilde T$ is the
normalized stress tensor satisfying $\lan \tilde T \tilde T\ran$ $\sim
O(1)$) from the AdS computation is now $\sim \sqrt{\k}$ which matches
with the field theory result $O(1/N)$.\footnote{We made these
  arguments for a large $N$ gauge theory such as ${\cal N}=4$ SYM, but
  for vector models and other examples, this counting can be
  appropriately modified.}  In the above we have assumed that the
scaling dimension of $\mO(x)$ is $O(1)$ (compared with $N$, or more
generally, with the central charge $c$ of the CFT). The back-reaction
on the metric is then given by the equation of motion for the graviton
$\del^2 h \sim \sqrt{\k} \lan \del \phi \del \phi \ran$. Now $\lan
\phi \phi \ran \sim O(1)$ since $\phi$ is canonically
normalized. (Alternatively, $\lan \phi \phi \ran$ is related to $\lan
\mO \mO \ran$ by bulk-boundary correspondence and the latter is, by
convention, $O(1)$. We could also arrive at this result by noting that
$\delta g \sim G_N T_{{\rm bulk}, \mu\nu}$ which is $\sim G_N \lan O|
T_{\mu\nu}| O \ran$ $\sim G_N \sim 1/N^2$ (which matches $h\sim
1/N$. From the last point of view, it is clear that we need the single
trace operator to have scaling dimension $\Delta \sim O(1)$.

The above argument about probe approximation can be easily extended to
the case when the CFT is deformed by both single trace and double
trace operators. The zero-th order bulk scalar action, $S_b$ remains
quadratic. 

We should make a remark here about self-interaction of the bulk
scalar. Typically the connected 3-point function $\lan \mO \mO \mO
\ran$ will be non-vanishing. But this will also be $O(1/N)$. Hence
$S_{int}$ will have a term $\sim \int \sqrt\k \phi^3$. 

\paragraph{Justification of Hamilton-Jacobi:} We argued above
that in the large $N$ approximation, it suffices to consider
a quadratic action, making Hamilton-Jacobi approximation to the
Schr\"{o}dinger equation is exact (up to a pre-factor which is
not important for our purpose).

\end{appendices}

{\normalsize
  \bibliographystyle{JHEP}
  \bibliography{v1a}

\providecommand{\href}[2]{#2}\begingroup\raggedright\begin{thebibliography}{10}

\bibitem{deHaro:2000vlm}
S.~de~Haro, S.~N. Solodukhin, and K.~Skenderis, {\it {Holographic
  reconstruction of space-time and renormalization in the AdS / CFT
  correspondence}},  {\em Commun. Math. Phys.} {\bf 217} (2001) 595--622,
  [\href{http://arxiv.org/abs/hep-th/0002230}{{\tt hep-th/0002230}}].

\bibitem{Heemskerk:2010hk}
I.~Heemskerk and J.~Polchinski, {\it {Holographic and Wilsonian Renormalization
  Groups}},  {\em JHEP} {\bf 1106} (2011) 031,
  [\href{http://arxiv.org/abs/1010.1264}{{\tt arXiv:1010.1264}}].

\bibitem{Faulkner:2010jy}
T.~Faulkner, H.~Liu, and M.~Rangamani, {\it {Integrating out geometry:
  Holographic Wilsonian RG and the membrane paradigm}},  {\em JHEP} {\bf 1108}
  (2011) 051, [\href{http://arxiv.org/abs/1010.4036}{{\tt arXiv:1010.4036}}].

\bibitem{Gubser:1998bc}
S.~S. Gubser, I.~R. Klebanov, and A.~M. Polyakov, {\it {Gauge theory
  correlators from noncritical string theory}},  {\em Phys. Lett.} {\bf B428}
  (1998) 105--114, [\href{http://arxiv.org/abs/hep-th/9802109}{{\tt
  hep-th/9802109}}].

\bibitem{Witten:1998qj}
E.~Witten, {\it {Anti-de Sitter space and holography}},  {\em Adv. Theor. Math.
  Phys.} {\bf 2} (1998) 253--291,
  [\href{http://arxiv.org/abs/hep-th/9802150}{{\tt hep-th/9802150}}].

\bibitem{Witten:2001ua}
E.~Witten, {\it {Multitrace operators, boundary conditions, and AdS / CFT
  correspondence}},  \href{http://arxiv.org/abs/hep-th/0112258}{{\tt
  hep-th/0112258}}.

\bibitem{Klebanov:1999tb}
I.~R. Klebanov and E.~Witten, {\it {AdS / CFT correspondence and symmetry
  breaking}},  {\em Nucl. Phys.} {\bf B556} (1999) 89--114,
  [\href{http://arxiv.org/abs/hep-th/9905104}{{\tt hep-th/9905104}}].

\bibitem{Balasubramanian:1999re}
V.~Balasubramanian and P.~Kraus, {\it {A Stress tensor for Anti-de Sitter
  gravity}},  {\em Commun. Math. Phys.} {\bf 208} (1999) 413--428,
  [\href{http://arxiv.org/abs/hep-th/9902121}{{\tt hep-th/9902121}}].

\bibitem{MN:2016mathematica}
{\em {\bf Mathematica notebooks supplied as `Ancillary files' along with the
  present ArXiv submission:} {\small CalculationsFile.nb}}.

\bibitem{Elander:2011vh}
D.~Elander, H.~Isono, and G.~Mandal, {\it {Holographic Wilsonian flows and
  emergent fermions in extremal charged black holes}},  {\em JHEP} {\bf 1111}
  (2011) 155, [\href{http://arxiv.org/abs/1109.3366}{{\tt arXiv:1109.3366}}].

\bibitem{Aharony:2015afa}
O.~Aharony, G.~Gur-Ari, and N.~Klinghoffer, {\it {The Holographic Dictionary
  for Beta Functions of Multi-trace Coupling Constants}},  {\em JHEP} {\bf 05}
  (2015) 031, [\href{http://arxiv.org/abs/1501.06664}{{\tt arXiv:1501.06664}}].

\bibitem{Pomoni:2008de}
E.~Pomoni and L.~Rastelli, {\it {Large N Field Theory and AdS Tachyons}},  {\em
  JHEP} {\bf 04} (2009) 020, [\href{http://arxiv.org/abs/0805.2261}{{\tt
  arXiv:0805.2261}}].

\bibitem{Balasubramanian:1999zv}
V.~Balasubramanian and S.~F. Ross, {\it {Holographic particle detection}},
  {\em Phys. Rev.} {\bf D61} (2000) 044007,
  [\href{http://arxiv.org/abs/hep-th/9906226}{{\tt hep-th/9906226}}].

\bibitem{Louko:2000tp}
J.~Louko, D.~Marolf, and S.~F. Ross, {\it {On geodesic propagators and black
  hole holography}},  {\em Phys. Rev.} {\bf D62} (2000) 044041,
  [\href{http://arxiv.org/abs/hep-th/0002111}{{\tt hep-th/0002111}}].

\bibitem{deBoer:1999tgo}
J.~de~Boer, E.~P. Verlinde, and H.~L. Verlinde, {\it {On the holographic
  renormalization group}},  {\em JHEP} {\bf 08} (2000) 003,
  [\href{http://arxiv.org/abs/hep-th/9912012}{{\tt hep-th/9912012}}].

\bibitem{Verlinde:1999xm}
E.~P. Verlinde and H.~L. Verlinde, {\it {RG flow, gravity and the cosmological
  constant}},  {\em JHEP} {\bf 05} (2000) 034,
  [\href{http://arxiv.org/abs/hep-th/9912018}{{\tt hep-th/9912018}}].

\bibitem{Balasubramanian:1999jd}
V.~Balasubramanian and P.~Kraus, {\it {Space-time and the holographic
  renormalization group}},  {\em Phys. Rev. Lett.} {\bf 83} (1999) 3605--3608,
  [\href{http://arxiv.org/abs/hep-th/9903190}{{\tt hep-th/9903190}}].

\bibitem{deBoer:2000cz}
J.~de~Boer, {\it {The Holographic renormalization group}},  {\em Fortsch.
  Phys.} {\bf 49} (2001) 339--358,
  [\href{http://arxiv.org/abs/hep-th/0101026}{{\tt hep-th/0101026}}].

\bibitem{Akhmedov:2002gq}
E.~T. Akhmedov, {\it {Notes on multitrace operators and holographic
  renormalization group}},  in {\em {Workshop on Integrable Models, Strings and
  Quantum Gravity Chennai, India, January 15-19, 2002}}, 2002.
\newblock \href{http://arxiv.org/abs/hep-th/0202055}{{\tt hep-th/0202055}}.

\bibitem{Laia:2011wf}
J.~N. Laia and D.~Tong, {\it {Flowing Between Fermionic Fixed Points}},  {\em
  JHEP} {\bf 11} (2011) 131, [\href{http://arxiv.org/abs/1108.2216}{{\tt
  arXiv:1108.2216}}].

\bibitem{Grozdanov:2011aa}
S.~Grozdanov, {\it {Wilsonian Renormalisation and the Exact Cut-Off Scale from
  Holographic Duality}},  {\em JHEP} {\bf 06} (2012) 079,
  [\href{http://arxiv.org/abs/1112.3356}{{\tt arXiv:1112.3356}}].

\bibitem{Balasubramanian:2012hb}
V.~Balasubramanian, M.~Guica, and A.~Lawrence, {\it {Holographic
  Interpretations of the Renormalization Group}},  {\em JHEP} {\bf 01} (2013)
  115, [\href{http://arxiv.org/abs/1211.1729}{{\tt arXiv:1211.1729}}].

\bibitem{Susskind:1998dq}
L.~Susskind and E.~Witten, {\it {The Holographic bound in anti-de Sitter
  space}},  \href{http://arxiv.org/abs/hep-th/9805114}{{\tt hep-th/9805114}}.

\bibitem{Vecchi:2010jz}
L.~Vecchi, {\it {The Conformal Window of deformed CFT's in the planar limit}},
  {\em Phys.Rev.} {\bf D82} (2010) 045013,
  [\href{http://arxiv.org/abs/1004.2063}{{\tt arXiv:1004.2063}}].

\bibitem{Vecchi:2010dd}
L.~Vecchi, {\it {Multitrace deformations, Gamow states, and Stability of
  AdS/CFT}},  {\em JHEP} {\bf 04} (2011) 056,
  [\href{http://arxiv.org/abs/1005.4921}{{\tt arXiv:1005.4921}}].

\bibitem{vanRees:2011fr}
B.~C. van Rees, {\it {Holographic renormalization for irrelevant operators and
  multi-trace counterterms}},  {\em JHEP} {\bf 08} (2011) 093,
  [\href{http://arxiv.org/abs/1102.2239}{{\tt arXiv:1102.2239}}].

\bibitem{Moshe:2003xn}
M.~Moshe and J.~Zinn-Justin, {\it {Quantum field theory in the large N limit: A
  Review}},  {\em Phys. Rept.} {\bf 385} (2003) 69--228,
  [\href{http://arxiv.org/abs/hep-th/0306133}{{\tt hep-th/0306133}}].

\end{thebibliography}\endgroup
}

\end{document}